\documentclass[12pt,preprint]{aastex}

\shorttitle{Superflare on EV Lac}
\shortauthors{Osten et al.}

\begin{document}

\newcommand{\Ka}{K$\alpha$ }

\title{ The Mouse that Roared:  A Superflare from the dMe Flare Star EV Lac
detected by Swift and Konus-Wind}
\author{Rachel A. Osten}
\affil{Space Telescope Science Institute, 3700 San Martin Drive, Baltimore MD 21218 \\ osten@stsci.edu}

\author{Olivier Godet}
\affil{Universit\'{e} de Toulouse, UPS, CESR, 
9 avenue du Colonel Roche, 
31028 Toulouse Cedex 9, France  \\Olivier.Godet@cesr.fr}

\author{Stephen Drake\altaffilmark{2}, Jack Tueller, Jay Cummings, Hans Krimm}
\affil{NASA Goddard Space Flight Center, Greenbelt, MD 20771}
\altaffiltext{2}{Also USRA}

\author{John Pye}
\affil{Department of Physics \& Astronomy, University of Leicester,
University Road, Leicester LE1 7RH UK}

\author{Valentin Pal'shin, Sergei Golenetskii}
\affil{Ioffe Physico-Technical Institute\\
Laboratory for Experimental Astrophysics\\
26 Polytekhnicheskaya, St Petersburg 194021,\\
Russian Federation}

\author{Fabio Reale}
\affil{Dip. Scienze Fis. \& Astron., Sez. Astron.,Universit{\'{a}} di Palermo,P.za Parlamento 1, 90134 Palermo, Italy}

\author{Samantha R. Oates, Mat J. Page}
\affil{Mullard Space Science Laboratory, University College London, Holmbury St. Mary, Dorking, Surrey, RH5 6NT, UK}

\author{Andrea Melandri}
\affil{Astrophysics Research Institute, Liverpool John Moores University,
Twelve Quays House, Birkenhead CH41 1LD UK}

\begin{abstract}
We report on a large stellar flare from the nearby dMe flare star EV Lac  observed by the Swift and Konus-Wind
satellites and the Liverpool Telescope. 
It is the first large stellar flare from a dMe flare star to result in a Swift trigger based on its hard X-ray
intensity. 
Its 
peak f$_{X}$ from 0.3--100 keV of 5.3$\times$10$^{-8}$ erg cm$^{-2}$ s$^{-1}$ is nearly 7000 times
larger than the star's quiescent coronal flux,
and the change in magnitude in the white filter is $\geq$4.7.
This flare also 
caused a transient increase in 
EV Lac's bolometric luminosity (L$_{\rm bol}$) during the early stages of the flare, with a peak estimated
L$_{X}$/L$_{\rm bol}$ $\sim$3.1.
We apply
flare loop hydrodynamic
modeling to the plasma parameter temporal changes to derive a loop semi-length of $l/R_{\star}=$ 0.37$\pm$0.07.
The soft X-ray spectrum of
the flare reveals evidence of iron \Ka emission at 6.4 keV.
We model the \Ka emission as fluorescence 
from the hot flare source irradiating the photospheric iron, and derive loop heights of $h/R_{\star}$=0.1,
consistent within factors of a few with the heights inferred from hydrodynamic modeling.
The \Ka emission feature shows variability on time scales of $\sim$200 s
which is difficult to interpret using the pure fluorescence hypothesis.  
We examine \Ka emission produced by collisional ionization from accelerated particles,
and find parameter values for the spectrum of accelerated particles which can accommodate the increased amount of \Ka flux
and the lack of observed nonthermal emission in the 20-50 keV spectral region. 
\end{abstract}

\keywords{stars: activity --- stars: coronae --- stars: flare --- stars: late-type --- 
stars: individual (EV~Lac) --- X-rays: stars -- Swift
}

\section{Introduction}
The origin of solar and stellar flares is thought to be the impulsive release of energy from
a non-potential magnetic field due to an instability arising from magnetic loop foot-point motions in the photosphere, 
resulting in plasma heating, particle acceleration, shocks and mass motions,
involving the entire outer atmosphere, and producing emissions across the electromagnetic spectrum.
Flares of all sizes 
are useful probes of coronal structures, under the assumption that there
is no difference in the flaring and non-flaring coronal structures during the
impulsive and transitory energy input.
Extreme flare events are useful for revealing the extent to which solar
flare physics can be extrapolated (and thus the applicability of the solar model in a variety of
stellar environments), and 
for enabling stellar flare diagnostic techniques which require high signal-to-noise.  

It has long been known that the largest observed stellar flares on ``normal'' stars (i.e. non-degenerate,
non-accreting stars)
represent energetic 
releases which are orders of magnitude more than typical solar flares  --- up to
a factor of 10$^{6}$ more energetic
\citep[compared to a typical solar flare energy of $10^{32}$ erg;][]{emslie2004}, i.e. $\sim$ 10$^{38}$
erg \citep{ks1996}.  
Yet the energy radiated as a result of plasma heating usually represents only a small fraction
of the total energy involved in a solar flare: \citet{emslie2004} estimate that 
the energy contribution of nonthermal particles in solar flares is $\approx$ a factor of 10 higher than
the amount of thermal plasma heating as expressed in their soft X-ray emission, and \citet{woods2004,woods2006} showed that
the total solar flare energy can be up to 110 times that seen in the GOES soft X-ray band from 1.2--12 keV.

White light solar flares are difficult to detect in total solar irradiance measurements
due to the contrast between the flaring emission and total solar disk
emission, whereas the red nature of the underlying M dwarf spectrum makes the blue flare continuum 
emission easier to detect.
In solar white light flares there is a timing and spatial correlation between the white light flare emission and
signatures of nonthermal emission such as hard X-ray emission \citep{neidigkane1993}, and radiative hydrodynamical simulations of
the impact of electron beams in the lower atmosphere can reproduce the observed solar continuum enhancements 
\citep{allred2005}, but extension of the same models to flares on M dwarfs has not to date been able to 
reproduce the observed
continuum enhancements \citep{allred2006}.
Large flares from M dwarf stars have been known at optical wavelengths for several decades; the
UV/optical/infrared spectral energy distribution typically peaks at short wavelengths, with
a roughly blackbody shape \citep{hawleyfisher1992} having a blackbody temperature $T_{\rm BB}$ of $\sim$10$^{4}$K.  
Enhancements of greater than a few magnitudes are large amplitude events which occur rarely;
\citet{welsh2007} found 7 out of 52 flares in the GALEX data archives to have $\Delta m_{\rm UV} \ge$ 4.4.
The most extreme stellar flare ever observed at ground-based near-UV/optical wavelengths may be that of V374 Peg 
\citep{dm11flare}, at $\Delta$U=11.0.

Soft X-ray flares result from the transient plasma heating that occurs as a result of energy deposition
lower in the atmosphere.  In the corona, magnetic field pressure exceeds the gas pressure
and therefore  dominates the dynamics.  The non-flaring coronal emission is easy to understand
in terms of an optically thin, multi-temperature, and possibly multi-density, plasma in coronal equilibrium. 
The X-ray spectral region also contains diagnostics which have been used recently in large stellar flares
to diagnose the presence of
nonthermal electrons operating during a large stellar flare \citep{osten2007} as well as 
probe the interaction between the corona and the lower atmosphere \citep{osten2007,testa2008}.
Changes in the characteristics of
the flaring plasma with time lead to deductions about heating time-scales, place constraints
on spatial scales, and provide evidence for temporary enrichment of coronal material with 
plasma from the lower layers of the atmosphere.
``Superflares'', in which the soft X-ray luminosity increases by $\gtrsim$ a factor
of 100 compared to its normal value, are rare but have been seen on several different active stars 
\citep{favata1999,franciosini2001,maggio2000,favata2000,osten2007}.
As the quiescent X-ray luminosity
of active stars is already at a high level relative to the underlying bolometric luminosity (L$_{X}$/L$_{\rm bol}$ $\approx$
10$^{-3}$), such an increase during a flare produces a temporary but significant
increase in the total luminosity of the star.


The nearby (d=5.06 pc) flare star EV~Lac has a record of frequent and large outbursts
across the electromagnetic spectrum.
\citet{kodaira1976} reported a long-lasting
($\Delta t\ge 3$ hrs) flare with a peak enhancement of $\Delta$U= 5.9, while
\citet{rs1982} observed a flare from EV~Lac with $\Delta$U= 6.4 at peak which lasted $\sim$ 9--11 hours.  
\citet{osten2005} noted frequent soft X-ray flares from EV~Lac.
Its bolometric luminosity L$_{\rm bol}$ has been estimated by \citet{favata2000} as 5.25$\times$10$^{31}$ erg s$^{-1}$.
Large-scale magnetic fields covering a substantial fraction of the surface ($B$ $\sim$ 4 kG, $f$ $\sim$50\%)
have been measured \citep{cmj1996}.
The extreme magnetic activity as revealed by
the flare observations is mirrored in complementary measures of changes in the large-scale
magnetic field configuration on relatively short time-scales of less than an hour \citep{pb2006}.

Here we describe the properties of a flare serendipitously observed on EV~Lac
by the Swift Gamma-ray Burst
satellite, the Konus GRB experiment on the Wind satellite, and observed in a fast follow-up by
the Liverpool Telescope.  
The flare is remarkable in the magnitude of the 
peak enhancement at optical and soft and hard X-ray energies.  The current paper provides
an overview of the datasets involved and their reduction (\S 2), determinations of variations in plasma parameters
during the flare, accompanying analysis of X-ray spectral features, and derivation of 
parameters from optical data (\S 3).  Section 4 discusses flare energy estimates, compares
the current flare with other large-scale flares, and the use of the Fe K$\alpha$ 6.4 keV
line to diagnose conditions in the flaring plasma, while Section 5 summarizes our conclusions.

%
\section{Data Sets}
An overview of the telescopes and observatories which obtained data for this EV~Lac flare is
given in Figure~\ref{fig:flarelc}, which delineates the timing and relative intensity of
the several X-ray and $\gamma$-ray observations made by Swift and Konus, and also the timing of the
UV and optical observations made by Swift and on the ground from the Liverpool Telescope.
All timing is referenced to the Burst Alert Telescope (BAT)  trigger time, 2008-04-25 05:13:57 UT;
we refer to this as T0. 

\subsection{Konus-Wind}
Konus \citep{kwref}
is an instrument on the Wind spacecraft which provides omni-directional and continuous
coverage of cosmic gamma-ray transients.  The flare was detected in waiting mode
by the S2 detector which observes the north ecliptic hemisphere.
The background is very stable in the S2 detector.
There is a statistically significant bump in count rate in the lowest energy range (18-70 keV) 
from $\sim$T0-100:T0+140 s of 5875$\pm$692 counts, which
likely indicates the beginning of the superflare. 
There is also a possible tail until $\sim$T0+1000 s. 
In addition,  there may be
a precursor event from $\sim$T0-700:T0-400 s  with 2117$\pm$791 counts.
Timing analysis shows that the propagation time from the Wind spacecraft
to the Swift spacecraft is 2.125 s, so the BAT trigger time (UT 05:13:57) corrected for time-of-flight is UT 05:13:54.875
at Konus.  
The peak in the Konus light curve shown in Figure~\ref{fig:flarelc} indicates that
the flare peaked $\sim$60 seconds before the BAT trigger.
The Konus data can be used to estimate the onset and rise time of the hard X-ray emission from the flare, and indicate 
a relatively fast rise time of less than 2 minutes.  

\subsection{Swift/BAT Data }
\begin{figure}[htbp]
\begin{center}
\includegraphics[scale=0.35]{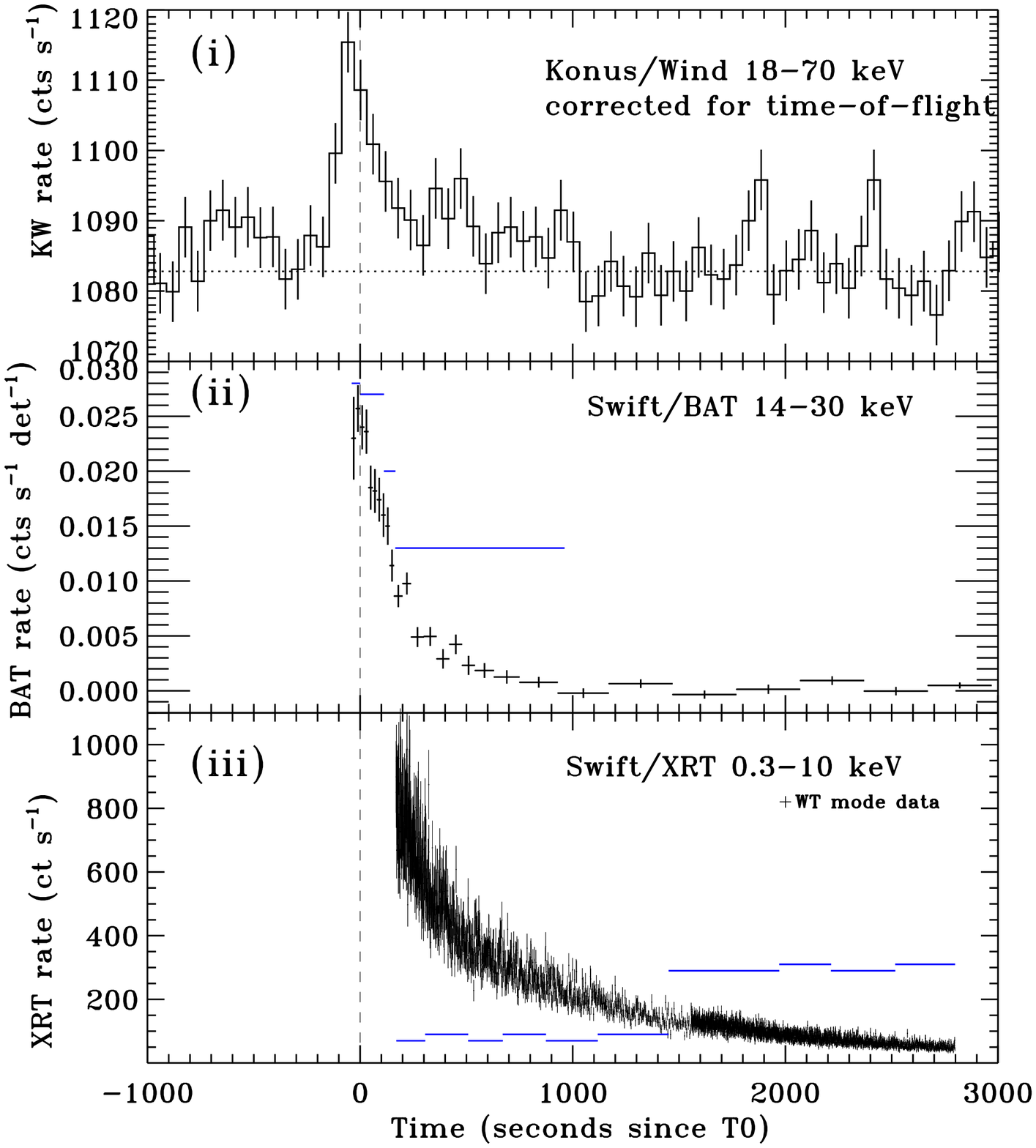}
\includegraphics[scale=0.35]{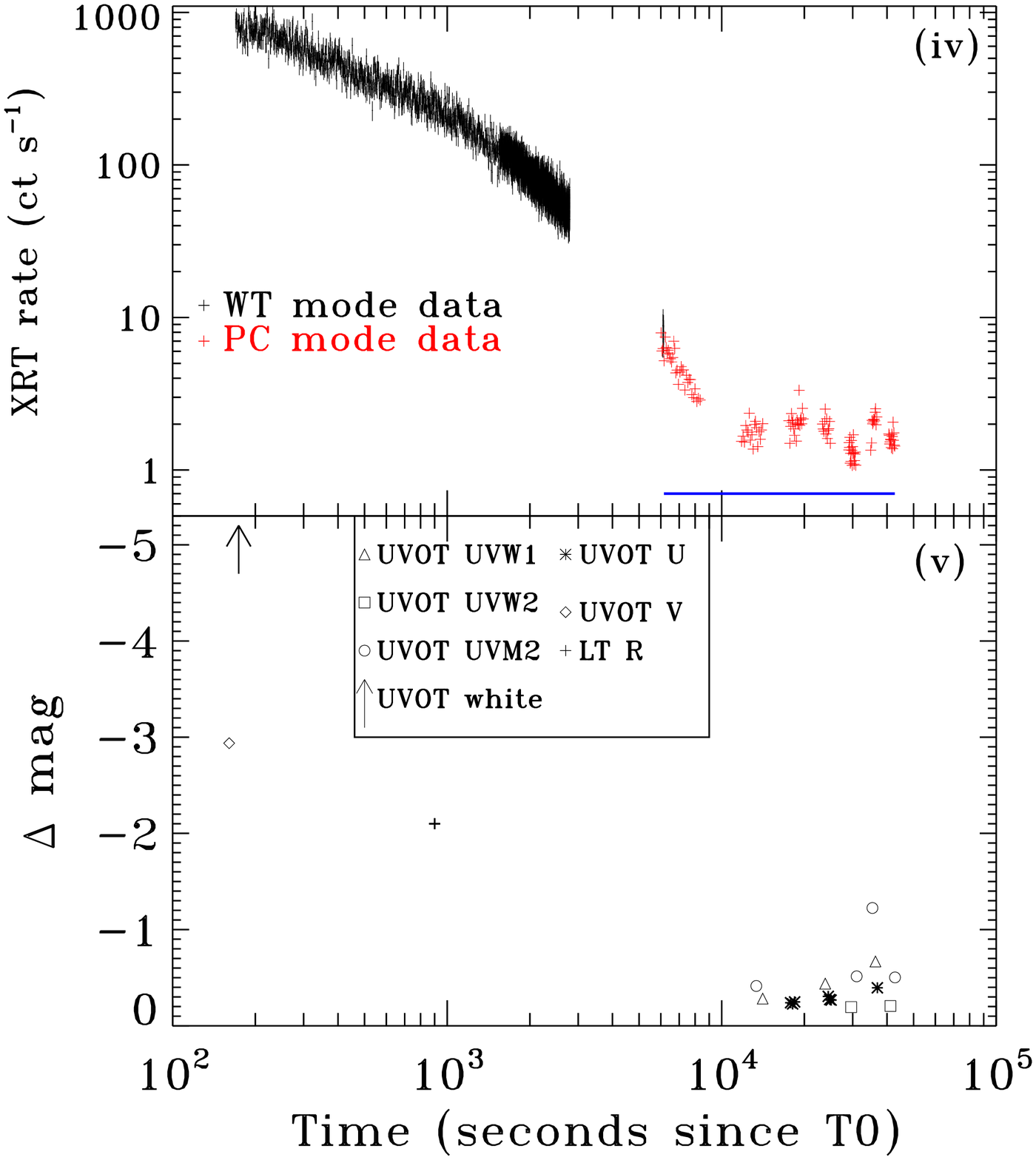}
\caption{Light curves during the flare from EV Lac.  
Panels $i$-$iii$ display emissions up to 3000 s past the trigger time, T0 of 2008-04-25T05:13:57,
depicting from top to bottom, the flare response as measured by Konus/Wind
in the 18-70 keV range ($i$), Swift/BAT from 14-30.1 keV ($ii$), Swift/XRT from 0.3-10 keV ($iii$).
Vertical dashed line indicates the BAT trigger time.
The Konus/Wind data have been binned to 58.88 s intervals, and have
been corrected for the time-of-flight between the Konus/Wind and Swift satellites: the horizontal
dotted line indicates the background
levels. Blue lines in the Swift/BAT and Swift/XRT panels indicate
the times over which spectra were extracted.
Panels $iv$ and $v$ display emissions up to 4$\times$10$^{4}$ s past T0.
For the XRT light curve ($iv$), black points indicate
data taken in Windowed Timing (WT) mode, red points indicate data taken in Photon Counting (PC) mode. 
The bottom panel ($v$) shows the UV and optical photometry for the flare.
\label{fig:flarelc}
}
\end{center}
\end{figure}

Swift's Burst Alert Telescope \citep{batref}
triggered on the flare from EV~Lac during a pre-planned spacecraft slew; the trigger
occurred
at 2008-04-25 UT 05:13:57 =T0. 
There was a short
pointing during which the discovery image was made with BAT, a second slew to place the star in the
apertures of the narrow-field instruments (X-Ray Telescope [XRT] and UV Optical Telescope [UVOT]), and a second longer pointing which overlaps the
XRT data.
The hard X-ray source was already declining in flux as it entered the BAT field of view during the pre-planned slew, with 
a peak intensity in the BAT 15-50 keV band of about 0.15 photons cm$^{-2}$ s$^{-1}$.
Comparison of the BAT and Konus/Wind light curves in Figure~\ref{fig:flarelc} shows that the
BAT triggered on the flare near its maximum in the hard X-ray band.
Light curves were extracted in the most sensitive 14--30.1 keV range where the spectral distribution
peaks, and were cleaned of the contaminating sources Cyg-X1, Cyg-X2, and 3A~0114+650 which were in
the partially coded field of view.
Spectra from 14-100 keV were extracted for four time intervals, indicated in Figure~\ref{fig:flarelc},
for which there was enough signal.

\subsection{Swift/XRT Data }
The X-Ray Telescope \citep{xrtref}
started to observe the source at 05:16:48\,UT (166\,s after the {\it
Swift}-BAT trigger, T0).
The XRT observes from 0.3-10 keV using CCD detectors, with energy resolution in the iron K region (6 keV)
of $\approx$ 140 eV as measured at launch time.
To produce the cleaned and calibrated event files, all the data were reduced
using the {\it Swift} {\scriptsize XRTPIPELINE} task (version 11.5) and
calibration files from the CALDB 2.8 release\,\footnote{See {\scriptsize
http://heasarc.gsfc.nasa.gov/docs/swift/analysis/}}.  A cleaned event list was
generated using the default pipeline, which removes the effects of hot pixels
and the bright Earth.  The initial data recording was in Windowed Timing (WT) mode, due to the large count
rate (initially $\approx$1000 count s$^{-1}$); from approximately 6160 s after the
trigger to the end of the observation (roughly 40ks after the BAT trigger)
data were taken in Photon Counting (PC) mode. 
The X-ray light curve was produced using the {\it Swift}-XRT
light curve generator \citep{evans2007}.
Note that this tool
takes into account the pile-up correction which is applied to the data when necessary.
From the cleaned event list, the source and background spectra were extracted
using {\scriptsize XSELECT}. We only use grade 0-2 events in WT mode and
grade 0-12 events in PC mode, which optimize the effective area and hence the
number of collected counts. The ancillary response files (ARF) for the PC and
WT modes were produced using {\scriptsize XRTMKARF}, taking into account the
point-spread function correction \citep{moretti2005}
as well as the exposure map correction. The exposure maps were
produced using the {\scriptsize XRTEXPOMAP} task (version 2.4). The v010 
response matrix files (RMFs)
({\scriptsize SWXWT0TO2S6$_{-}$20010101V010.RMF} in WT mode and {\scriptsize
SWXPC0TO12S6$_{-}$20010101V010.RMF} in PC mode) were used to perform the
spectral analysis \citep{godet2009}.

During the process of creation of the spectra and the light curve, we took 
great care to correct the data for the effect of pile-up in both PC and WT
modes. Indeed, the effects of pile-up result in an apparent loss of flux,
particularly from the center of the point spread function (PSF), and a 
migration of 0--12 grades (considered
as grades resulting from X-ray events) to higher grades and
energies at high count rates.  The high observed count rates in WT
mode (several hundred counts s$^{-1}$) and PC mode (above 2 counts
s$^{-1}$) indicate that the data are piled-up.
In WT mode, we produced a sample of grade ratio distribution using
background-subtracted source event lists created by increasing the size of the
inner radius of an annular extraction region with an outer radius of 30
pixels (1 pixel = 2.36"). The grade ratio distribution for grade 0, 1 \& 2 events is defined as
the ratio of grade 0, 1 \& 2 events over the sum of grade 0-2 events per
energy bin in the 0.3-10 keV energy range. We then compared this sample of
grade ratio distribution with that obtained using non piled-up WT data in
order to estimate the number of pixels to exclude, finding that an exclusion of
the innermost 6 pixels was necessary when all the WT data are used.
In PC mode, we used the same methodology as described in 
\citet{vaughan2006} 
to estimate the number of pixels which needs to be
excluded.
We fit the wings
of the source radial PSF profile with the XRT PSF model 
\citep[a King function; see][]{moretti2005}
and a background component
(here assumed to be a constant). The exclusion of the innermost 5 pixels
enables us to mitigate the effects of pile-up.

\subsection{Swift/UVOT Data }
The Ultra-Violet/Optical Telescope (UVOT; \cite{roming}) began
observing EV Lac, at UT 05:16:35 (157s after the {\it Swift}-BAT
trigger, T0), and performed a sequence of automatic
observations. The first observation was an 8.08s settling exposure,
which is usually taken in event/image mode with the $v$ filter while
the UVOT stabilizes and prepares for settled observations.  The second
observation, a 200s $white$ finding chart exposure, was not completed,
because the safety circuit tripped. The trip occurred at UT 05:16:51.3,
1.4s after the UVOT had finished moving the filter wheel to
$white$. The safety circuit tripped due to a rate of $\ge$300 000
count s$^{-1}$ being observed over 31 frames (each of 11ms), implying
that EV Lac must have been brighter than 7th magnitude in $white$ for
a minimum of 1.4s. 
The UVOT resumed observations at $\sim11.89$~ks,
taking multiple exposures using the 7 filters of the UVOT ($white$,
$v$, $b$, $u$, $uvw1$, $uvm2$, $uvw2$). Table~\ref{tbl:mag} lists
the observed magnitudes measured with UVOT at various times during the
flare.  
The errors are purely statistical and do not reflect any systematic or calibration uncertainties.
A description of the effective wavelengths and bandwidths of the 
Swift/UVOT filters can be found at \verb=http://swift.gsfc.nasa.gov/docs/swift/about_swift/uvot_desc.html=.

Emission in the $v$ band from EV Lac was saturated during the settling exposure, providing a limit
for the brightness in $v$ of $<$11.48 magnitude. However, during this
exposure, EV Lac was sufficiently bright that it was accompanied by an
out-of-focus halo of light that has undergone multiple reflections
within the instrument. Because the count rate of the halo and the flux of
the source are related linearly, we are able to use the halo to
determine the brightness of EV Lac during the settling exposure.

In order to provide a calibration for the halo of EV Lac, we searched
the {\it Swift} archive\footnote{http://heasarc.gsfc.nasa.gov/cgi-bin/W3Browse/swift.pl}
 for stars of known brightness,
which are bright enough to produce a halo. We identified 9 stars with
UVOT observations in the $v$ filter, with $v$ magnitudes in the GSC-2
database ranging from 6.07 to 10.51, and with suitable halos.
  
To determine the count rates of the halos in the UVOT images, source
regions and background regions were created for EV Lac and for each of
the 9 stars.  For each object, the source region was chosen to cover
the entire halo, but excluding the central object itself and its
diffraction spikes, any visible background sources, and the
diffraction troughs in the halo. The corresponding background region 
was taken as a set of circular apertures, placed on source-free areas
of the sky, with a total area of at least twice that of the source
region.
 
For each of the 9 stars, the count rate of the halo was determined and
then scaled to the source area of EV Lac. These values were then
plotted against the count rates that would be predicted in the absence
of coincidence loss in 5{\arcsec} radius apertures using the $v$
zero point given in \citet{poole}. A linear function was fitted to
obtain the ratio between the halo count rates and predicted count
rates of the stars.  We multiplied the halo count rate of EV Lac by
this ratio to obtain the corresponding 
coincidence-loss-free count rate of EV Lac in an
aperture of 5{\arcsec} radius, and converted this into a $v$ magnitude
using the zero point of \citet{poole}.  In order to determine the
calibration uncertainty on the $v$ magnitude, we divided the count
rates of the halo calibration stars by the corresponding values of the
best-fit linear model and then took the standard deviation of the
resulting numbers to be the 1$\sigma$ fractional error.
This leads to an estimate of $v_{\rm max}=$7.2$\pm$0.2, compared to a V of 10.1
during quiescence.

For the rest of the images, in order to be compatible with the UVOT
calibration \citep{poole}, source counts were obtained using an
aperture of radius of 5{\arcsec} and background counts were obtained
from a region of radius 20{\arcsec}, positioned on a source-free area
of the sky. EV Lac was found to be saturated in the $white$, $v$ and $b$
filters. The software used to determine the source count rates can be
found in the software release HEADAS 6.5 and version 20071106 of the
UVOT calibration files. Count rates were corrected for coincidence
loss and converted into magnitudes following the prescription in
\cite{poole}.


We used the Swift UVOT tool \footnote{http://www.mssl.ucl.ac.uk/www\_astro/uvot/uvot\_observing/uvot\_tool.html\\} 
to estimate the count rate and therefore
magnitude of the star during quiescence (for an M5 spectral type) 
in the $v$, $white$, and $u$ filters,
in order to determine the flare enhancements of the
star.
We obtained additional ToO observations of EV~Lac nearly a year later to determine its
quiescent magnitude in the $uvw1$, $uvw2$, and $uvm2$ filters.
These quiescent magnitudes are listed in Table\ref{tbl:mag} and
were subtracted from the flare magnitudes to investigate the
photometric variations at late times in the flare, as 
illustrated in the bottom panel of Figure~\ref{fig:flarelc}.

\subsection{Liverpool Telescope}
The Liverpool Telescope took a series of 3x10 s images in the SDSS-r filter $\sim$ 15 minutes
after the trigger as part of the standard GRB follow-up.  No further observations were
taken once the trigger was identified as a non-GRB.  Photometry was done with GAIA
calibrating SDSS-r images against the USNOB1-R2 magnitudes of the few stars
in the field.  The night was not photometric so systematic errors of 0.2 mag (1 $\sigma$) are
added to all measurements, which also accounts for the difference between the R2
magnitude of the USNOB1 catalog \citep{monet2003} and
the r' band.
Photometry was performed on each of the 3 frames and a co-added 30 s frame using an aperture with
a radius of 8 arcsec.  The differences between magnitudes in the three 10 s frames were $<$0.05 mag
so we decided to estimate the final magnitudes on the co-added frame.  This results in the R2 magnitude
of EV~Lac being 7.4$\pm$0.2. 
There is a nearby star, Gl~873B, which at the time of the observations was $\approx$0.5 arcminutes from EV~Lac, 
which was used
to check the photometry.  Its magnitude in the R2 band is measured to be 10.9$\pm$0.2, and its catalogued
magnitude is R2$\sim$10.6, consistent within the errors with the value obtained here.
The estimated magnitude for EV~Lac during this observation is $\sim$ 2 magnitudes brighter than the
catalogued one (R2$\sim$9.5).

\section{Analysis}
\subsection{BAT Spectral Analysis}
\begin{figure}
\begin{center}
\end{center}
\includegraphics[scale=0.7,angle=90]{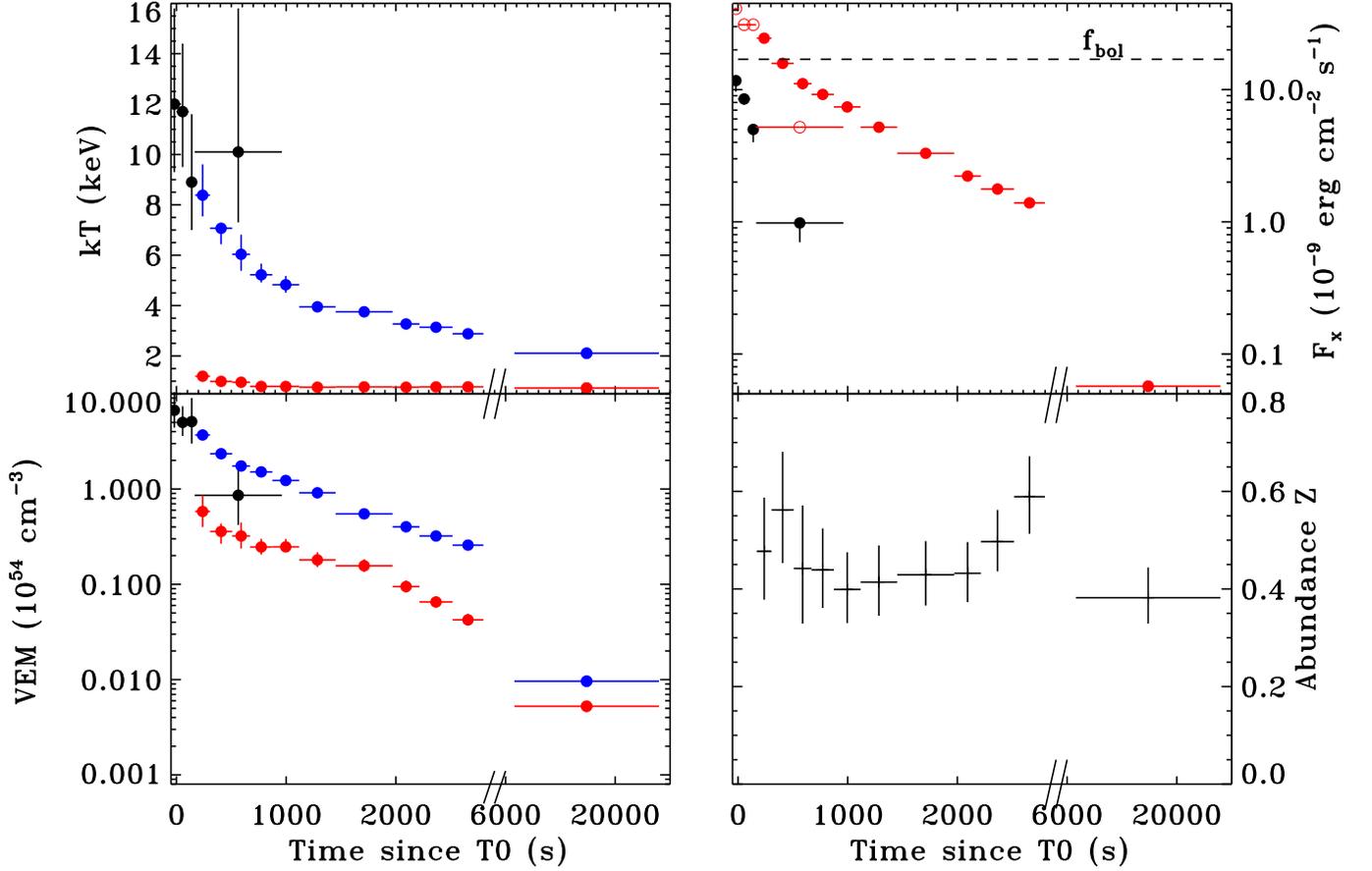}
\caption[]{ Results of time-resolved spectroscopic analysis.  Upper left panel
details variation of plasma temperature as derived from BAT spectra (black circles)
and the low and high temperatures determined from XRT spectra (red and blue circles,
respectively).  Horizontal bars give time range over which spectra were extracted;
vertical bars are 90\% error ranges.
Upper right panel shows variation of flux in the 14-100 keV range (black circles)
and 0.3--10 keV range (red circles).  Filled red circles correspond to measurements from
XRT spectra; open red circles are extrapolations from the best fit model to the BAT data.
Dashed line indicates bolometric flux of 1.7$\times$10$^{-8}$ erg cm$^{-2}$ s$^{-1}$.
Lower left panel shows trend in volume emission measure derived from APEC model fits
to BAT data (black circles), and the low and high temperature components fit to XRT
spectra (red and blue circles, respectively).
Lower right panel plots variation in metal abundance $Z$ as determined from spectral 
fitting to XRT data.
  }
\label{fig:specpar}
\end{figure}

We have performed a spectral analysis using XSPEC v12.4.0 \citep{xspecref} of the BAT data for several time intervals during which
the hard X-ray spectra were the only data collected: viz., during the first slew, 
first pointing, second slew.  During the second pointing, (from $\approx$ T0+166:T0+961 s) XRT
data were also collected.  The time intervals during which BAT spectra
were accumulated  are indicated in Figure~\ref{fig:flarelc} (b) 
with blue horizontal lines.  Spectral fitting of the 14-100 keV band used a single
temperature APEC model \citep{apecref} with abundances fixed to a multiple of the solar values of 
\citet{angr}. The \citet{angr} abundance table was used here for better comparison of
our results to other X-ray spectroscopic observations of EV Lac, e.g. \citet{evlacsuzaku} and \citet{osten2005}.
Because the standard APEC model included in the XSPEC distribution only considers emission
up to photon energies of 50 keV and the BAT energy range extends to 200 keV, we made use of
a custom APEC model calculated out to a maximum energy bin of 100 keV in order to model
correctly the bremsstrahlung contribution at high energies.
Based on the results from XRT fitting described below, we fixed the abundance in the APEC model 
used to fit the BAT data to $Z=0.4$ in order to facilitate comparison between BAT-only and XRT-only spectral fits,
the BAT data only being able to constrain the product of metallicity and volume emission measure.
Figure~\ref{fig:specpar} illustrates the variation in plasma temperature, flux, and
volume emission measure (VEM) derived from the BAT spectra, and Table~\ref{tbl:batspec}
lists the plasma parameters and errors derived from fitting BAT spectra.
The inferred plasma temperature during the first $\sim$100 seconds of BAT observations
is $\sim$12 keV.
To estimate the amount of soft X-ray flux during the times when BAT data 
were collected, we took the best-fit APEC model 
from fits to the BAT data and extrapolated to determine the 0.3-10 keV flux.
These are indicated in open red circles in the upper right panel of Figure~\ref{fig:specpar}.
The errors shown are 90\% confidence intervals, equivalent to $\pm$1.6$\sigma$.

\subsection{XRT Spectral Analysis \label{sec:xrtspec}}
\subsubsection{Time Variation of Plasma Parameters}
In order to investigate 
spectral evolution during the X-ray flare, we performed a detailed time-resolved
analysis in WT mode. The statistics in PC mode after pile-up correction are not
good enough to allow us to perform a meaningful time-resolved analysis. We
divided the WT data into ten time bins so that we generated in each of the time
bins a high statistical quality source spectrum with $2\times 10^4$ counts after
correcting for pile-up. The length of the time bins is variable, ranging from 
136--521 seconds.
Note that the inner radius of the annular extraction
region used to create the spectra decreases with time since the count rate
also decreases and the effects of pile-up become less and less dominant. 
All XRT spectra were binned to contain more than 20 counts bin$^{-1}$, and
were fit from 0.3 to 10 keV within XSPEC v12.4.0.
\citep{xspecref}.
We performed a single temperature fitting of the data using the APEC scaled solar abundance model,
as well as 
spectral fits using two discrete temperature ($2T$) APEC components.
The abundances of \citet{angr} are used here as well.
We neglected the absorption column,
since the real column towards this very nearby star is probably too low to be very well constrained in the
0.3-10 keV band. 
The time variations of the derived parameters (plasma temperature,
X-ray flux, volume emission measure, and global metal abundance) 
for the $2T$ fits are shown
in Figure~\ref{fig:specpar}.
Table~\ref{tbl:xrtspec} lists the plasma parameters and their errors derived
from fitting the XRT spectra.
 Although the errors on the derived abundances during
the time slices are large, there is no evidence for a significant change
in the global metallicity during the flare decay; for the present purposes we
take the abundance $Z$ to be $\sim$ 0.4.
The quiescent X-ray flux (0.5--7 keV) from Chandra observations is $\approx$ 5.8$\times$10$^{-12}$
erg cm$^{-2}$ s$^{-1}$ \citep{osten2005}, and so
the peak measured XRT flux is $\sim$ 4200 
 times this, and even when the flare has declined in flux
during the last time bin obtained in PC mode, the flux inferred is still  a factor
of 10 larger than the Chandra value.

\subsubsection{The Emission Line at 6.4 keV\label{sec:feka}}
The $2T$ APEC spectral fits to the WT mode XRT spectrum obtained from T0+171:T0+2798 s revealed
positive residuals around 6-7 keV which look like a line
feature not included in the APEC line list. 
To investigate that, we added a
Gaussian component to the 2-T APEC model. 
The resolution of the XRT at 6 keV was $\approx$ 140 eV at launch 
\citep{osborne2005}, but the CCD has experienced degraded resolution
since then due mainly to radiation damage.  The RMFs for WT mode data
do not contain proper account of this line broadening at high energy
due to lack of calibration information.  The width of the Gaussian
was fixed to a value of $\sigma =0.075$ keV based on evolution
in PC mode data of the line width from the onboard calibration $^{55}$Fe 5.9 keV line source
\citep[see discussion in][]{godet2009}.
The addition
of the Gaussian results in a fit improvement of $\Delta \chi^2 = 21$ for two
parameters and does not change the other fitting parameters. 
The line centroid ($E = 6.42\pm 0.04$ keV) may
be consistent with a fluorescence Fe K$\alpha$ line.  
Note that the Fe K$\alpha$ fluorescence line is really a doublet with
energy separation $\sim$12 eV, but this energy difference is small
compared with the energy resolution at the time of the observations,
and so our use of a single Gaussian with a width of 75 eV, broadened by
the instrumental resolution of $\sim$ 140 eV is sufficient.
The use of an F-test
gives a probability of $8.5\times 10^{-4}$ suggesting that the line is
real. However, \cite{protassov2002} have opined that the
F-test should not be used to assess the significance of a line.

\begin{figure}
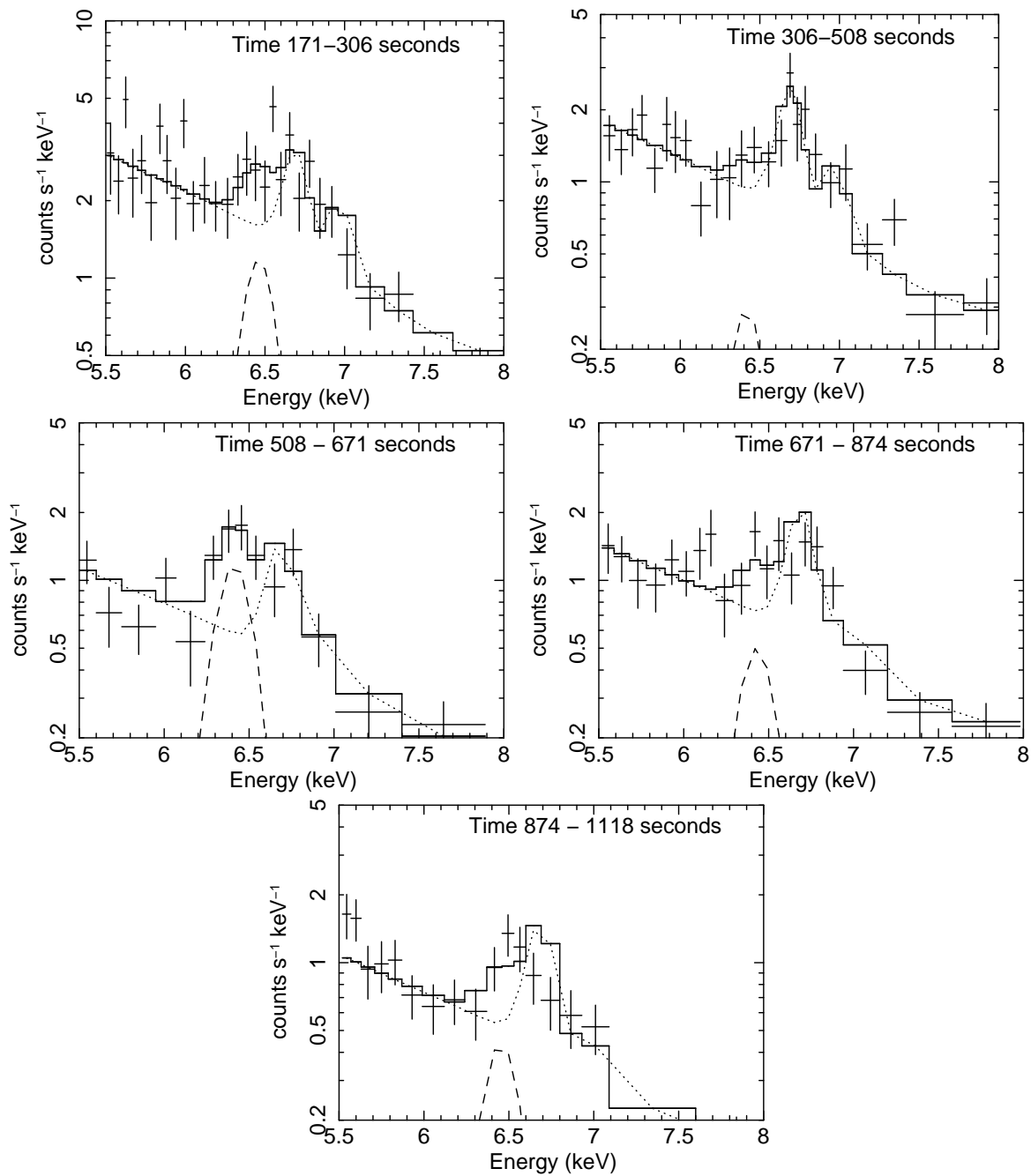

\begin{center}
\includegraphics[scale=0.3,angle=-90]{f3a.eps}
\includegraphics[scale=0.3,angle=-90]{f3b.eps}
\includegraphics[scale=0.3,angle=-90]{f3c.eps}
\includegraphics[scale=0.3,angle=-90]{f3d.eps}
\includegraphics[scale=0.3,angle=-90]{f3e.eps}
\caption[]{Close-up of the Fe~K region for the first five time intervals in
the initial stages of the flare decay.  Times are indicated in the top of the plot 
and refer to seconds since the BAT trigger time T0.  Figure~\ref{fig:flarelc}
shows the XRT light curve with lines indicating the times when these spectra were extracted.
Time intervals have been chosen so that spectra have a
minimum of 20,000 counts per spectrum before background subtraction.
Pluses are spectral data; dotted line is the best-fit two-temperature
APEC model; dashed line is the Gaussian fit to the feature at 6.4 keV,
and the solid histogram is the combination of the APEC and Gaussian models.
\label{fig:kaspec}
}
\end{center}
\end{figure}

To correctly work out the line significance, we used instead Monte-Carlo
simulations based on codes provided by C. Hurkett using the posterior
predictive $p$-value method. A detailed description of this method can be
found in Section 3.2 in \citet{hurkett2008}.
Here, we describe briefly
how the codes work.  The steps are: i) compute the
spectral parameters of the model used to fit the data;
ii) use these parameters with their errors to
generate $N$ sets of random parameter values and each of these is used to
generate a simulated spectrum; iii) fit the $N$ simulated spectra with the model
without a Gaussian line and create a table of the
$\chi^2$/d.o.f. values; iv) re-do the same as in iii) but including a Gaussian
line, the energy centroid varying around 6.4 keV (assuming that the feature is
due to a fluorescence Fe K$\alpha$ line) and the width being fixed at the
spectral resolution at the time of the observation ($\sigma = 0.075$ keV). A
new table of the $\chi^2$/d.o.f. values is created; v) compute the $\Delta
\chi^2$ distribution from the $N$ simulations and compare it with the $\Delta
\chi^2$ value obtained from the observed spectrum. The line significance is
then given by the number of times that the simulations produced a $\Delta
\chi^2$ value larger than the observed one.  In order to work, this method needs to run
a large number of simulations (here, we used $N=10^4$).  
We used the F-test probability to identify which time intervals 
should be explored further using this method.
When the F-test probability is low, the above method will not return a high probability,
but is needed in situations where the F-test gives a high probability; then the 
posterior predictive $p$-value method can determine whether the line is real.
Table~\ref{tbl:fek} lists the equivalent widths and fluxes of the 6.4 keV feature, along with
the results of the F-test and Monte Carlo simulations.
We performed Monte Carlo simulations for the first six time intervals; 
the F-test probability for time bins 7,8, and 10 were low so we did not 
do the additional Monte Carlo simulations for these intervals.  Time bin 9 
(T0+2215:T0+2517 s) shows the presence of a line with an F-test probability of
98.5\%, so we performed Monte Carlo simulations for this time interval.
The probability from Monte Carlo simulations is only 97.3\%, which we do not consider 
significant (we only considered probabilities $>$ 99\% to be statistically significant).
Figure~\ref{fig:kaspec} shows the \Ka region of the XRT spectrum
for the first five time intervals of the flare decay. 

The fit of the PC spectrum from $\sim$ T0+6.2$\times 10^3$:$\sim$ T0+
4.3$\times 10^4$\,s using a two temperature APEC model gives a
relatively good fit ($\chi^2/\mathrm{d.o.f.} = 236.3/175$).  
The addition of a
Gaussian line as in the WT spectrum does not improve the fit, not surprising
given the much lower flux in this dataset.
This indicates that the \Ka line is not present during this late stage of the flare decay.
The variability of the \Ka line strength is discussed further in \S 3.6.

\begin{figure}
\begin{center}
\includegraphics[scale=0.3]{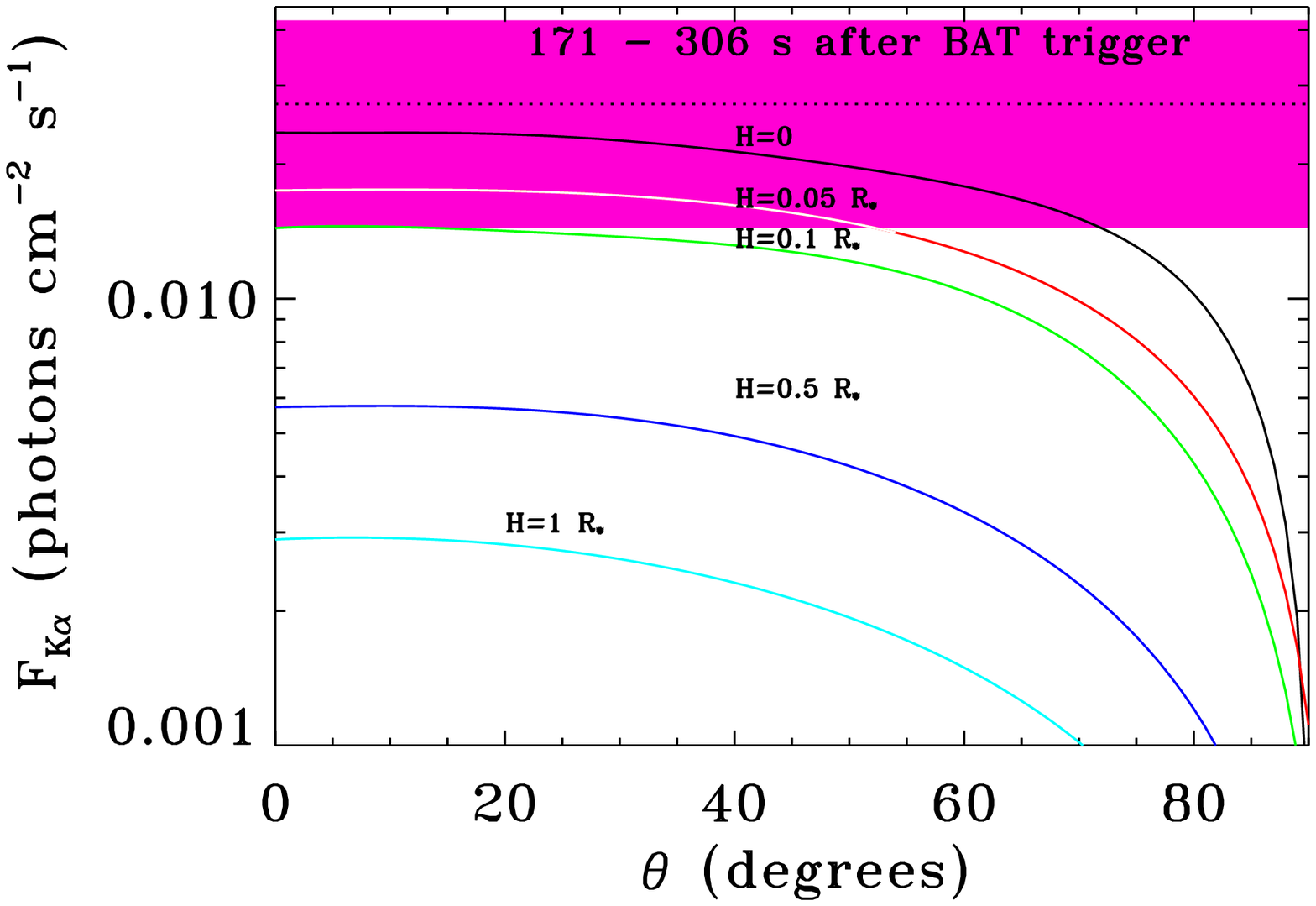}
\includegraphics[scale=0.3]{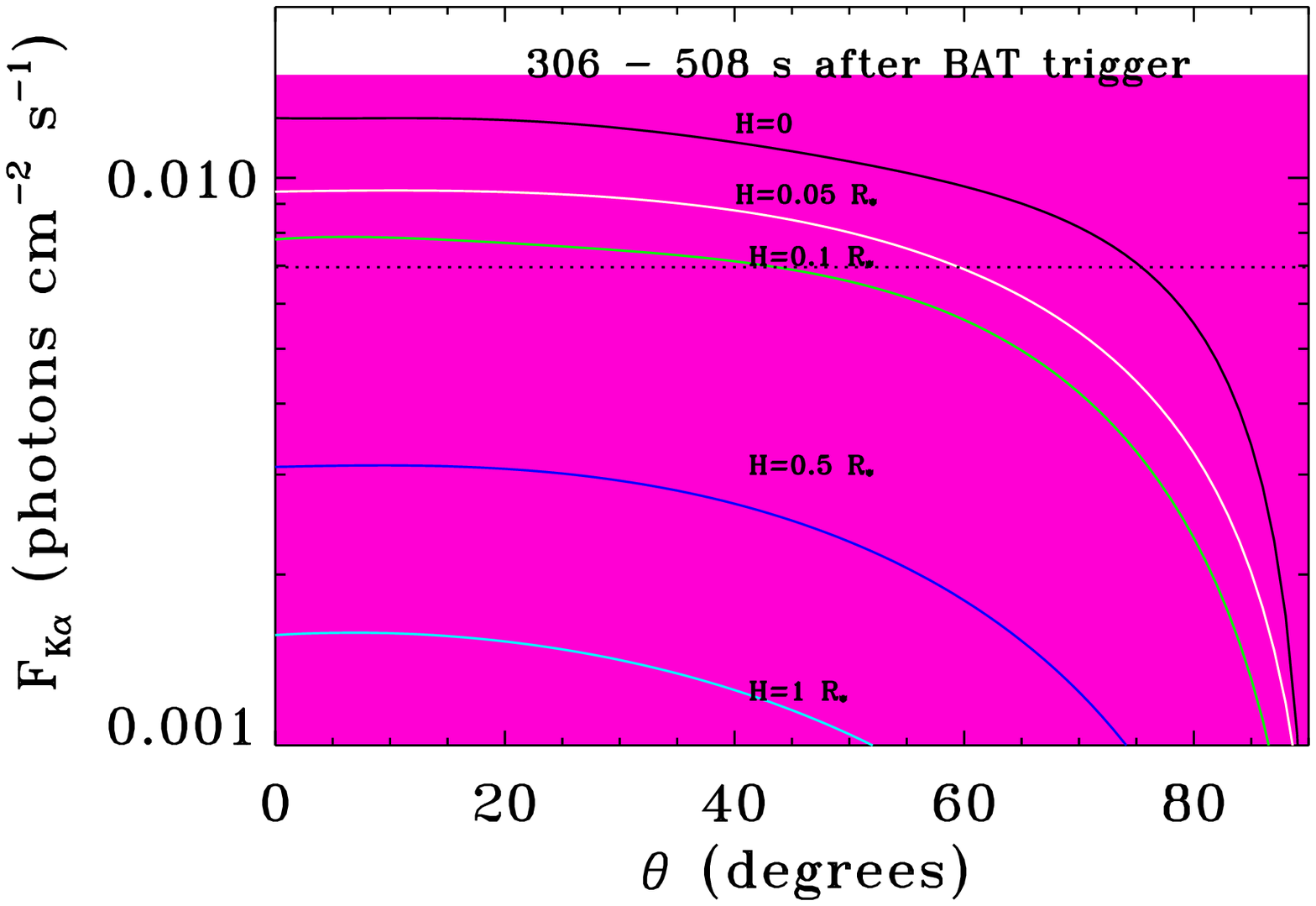}
\includegraphics[scale=0.3]{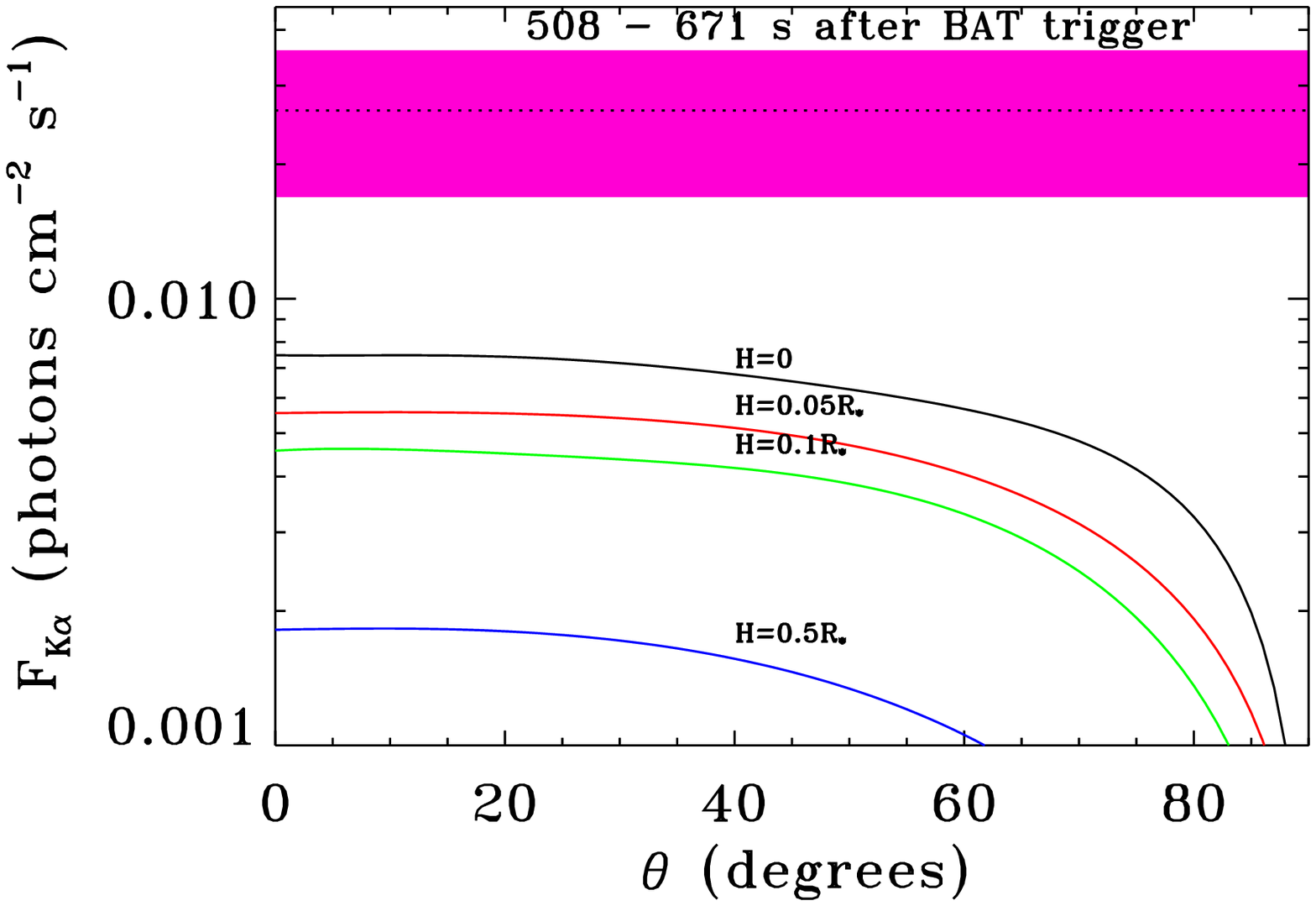}
\includegraphics[scale=0.3]{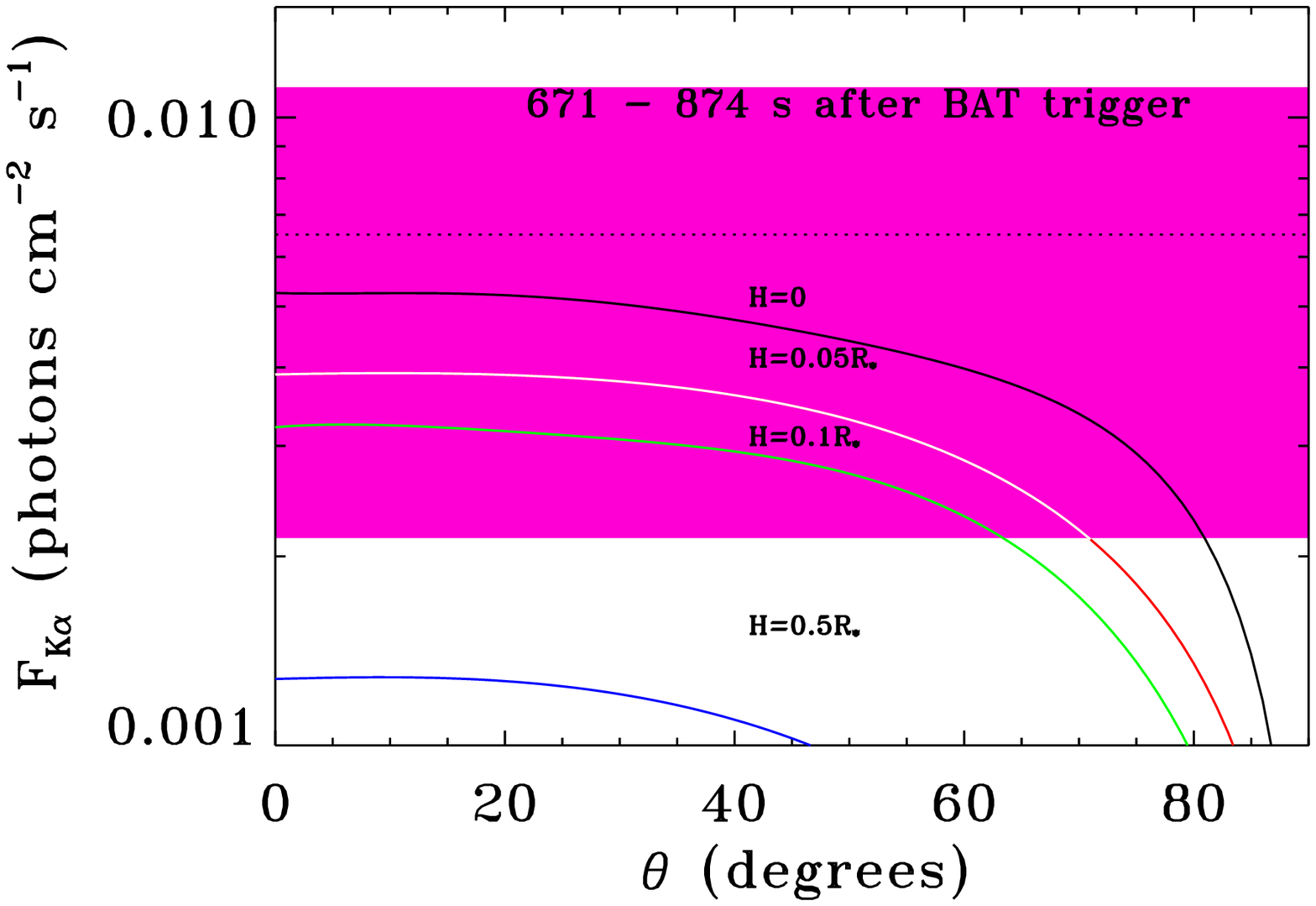}
\includegraphics[scale=0.3]{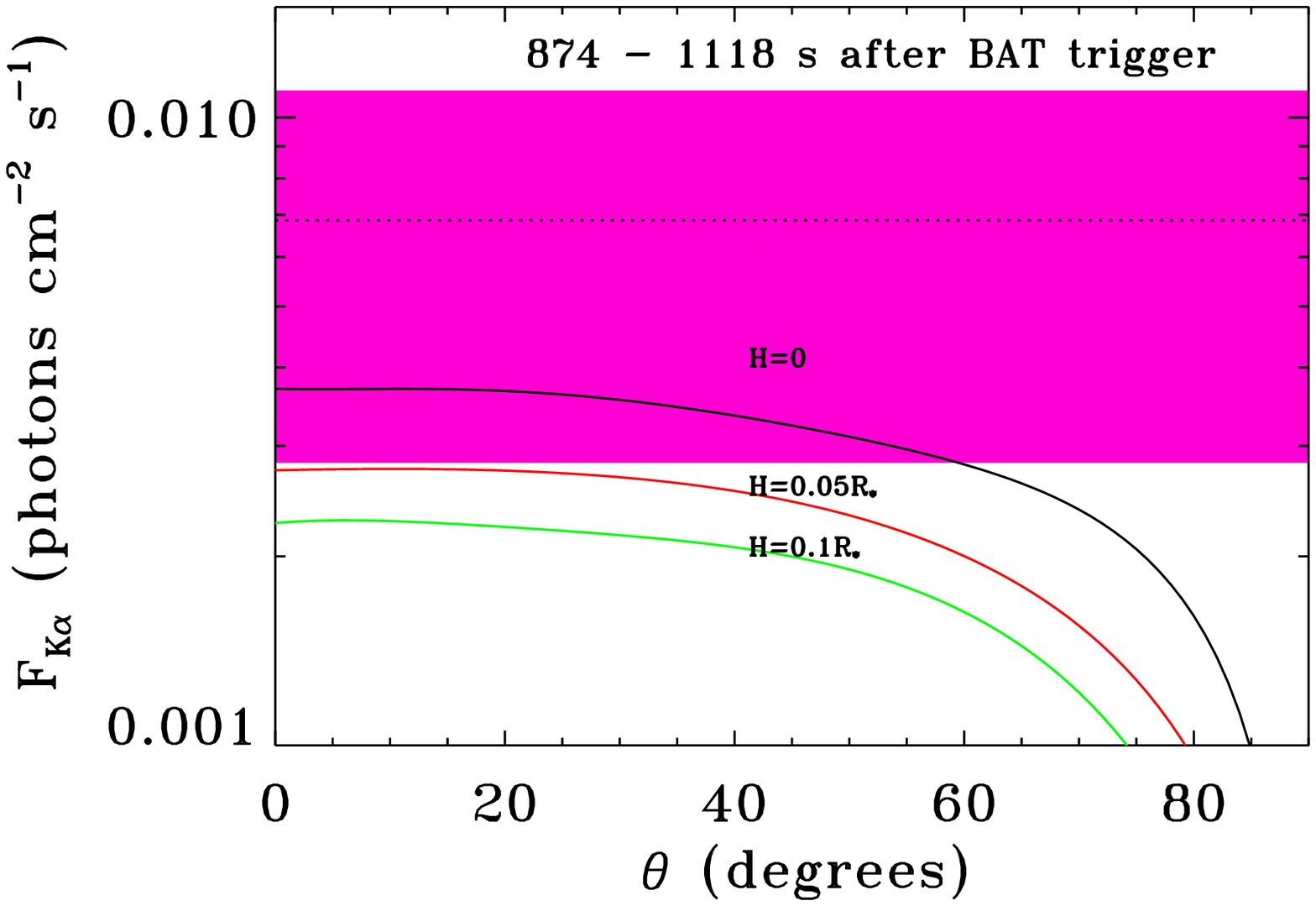}
\caption[]{Modelling of 6.4 keV line flux for the first five time intervals during the flare decay;
times are indicated corresponding to extracted spectra shown in Figure~\ref{fig:kaspec}.
Colored curves give expected run of K$\alpha$ fluorescent line flux as a function
of astrocentric angle for different flare loop heights.  Dotted line indicates 
measured \Ka flux and magenta region the 3$\sigma$ uncertainty in the \Ka flux.
\label{fig:kamodel}}
\end{center}
\end{figure}

The iron \Ka line is detected in a variety of X-ray emitting stars, from the solar corona
to compact stellar objects to actively accreting pre-main sequence stars
\citep{bai1979,gudelnaze2009}.
The formation is generally attributed to a fluorescent process, in which an X-ray
continuum source shines on a source of neutral or low ionization states of iron (\ion{Fe}{1}--\ion{Fe}{12}), 
photoionizing an inner K shell electron
and causing the de-excitation of an electron from a higher level at this energy.
The utility of this line stems from the 
dependence of the line strength on the 
solid angle subtended by the source of neutral or near-neutral iron. 
In the stellar context where the X-ray continuum emission arises from a loop-top source, and
the fluorescing material is photospheric iron,
the detection of this line thus allows for an independent constraint of the
height of the flaring loop above the photosphere.
The main contributors
to the observed flux in the \Ka line are the total photon luminosity above the \Ka ionization threshold
of 7.11 keV, the fluorescent efficiency of this radiation, and the photospheric iron abundance.
The fluorescent efficiency changes with photon energy due to the energy dependence of
the absorption cross section, and so the spectral distribution of photons, characterized
by the plasma temperature T of the X-ray-emitting material, is important for describing the fluorescence process.
The fluorescent efficiency also 
depends on the height of the X-ray continuum source above the fluorescent source (this arises
from the solid angle dependence).
The observed flux also depends on the astrocentric angle 
\citep[termed heliocentric angle in][]{drake2008,bai1979}; 
the flux is a maximum at zero, which corresponds to the center
of the stellar disk, and is a minimum when the source is located on the stellar limb 
(astrocentric angle of 90$^\circ$).
The observed flux is written as \\
\begin{equation}
F_{K\alpha} = \frac{f(\theta) \Gamma(T,h) N_{7.11}}{4\pi d^{2}} \;\;\; photons \; cm^{-2} \; s^{-1}
\end{equation}
where $N_{7.11}$ is the luminosity above 7.11 keV, $f(\theta)$ is a function
which describes the angular dependence of the emitted flux on astrocentric angle,
$\Gamma$ is the fluorescent efficiency, and $d$ is the distance to the object.  
This is equation 4 in \citet{drake2008}.
We use the values in Tables 2 and 3 of \citet{drake2008} for the dependence of $\Gamma$ on temperature
and height, and a parameterization of $f(\theta)$ for different heights \footnote{There is a typo in their
Table 3 for $h=0.5$R$_{\star}$ -- the $\alpha_{5}$ coefficient should
be $-6.63\times10^{–11}$.}.
\citet{drake2008} perform calculations using an iron abundance of $Fe/H=$3.16$\times$10$^{-5}$.
As we do not have a constraint on EV~Lac's photospheric iron abundance, we use this default value.
We use the best fit APEC model parameters for each time interval and create
a dummy response from 7--50 keV to compute the photon density spectrum above 7.11 keV,
$N_{7.11}/(4\pi d^{2})$.  
We use the $\Gamma$ calculated for a plasma
temperature of 100 MK ($\log T=8.0$), as this is closest to the temperatures determined from spectral fitting.
Figure~\ref{fig:kamodel} depicts the observed \Ka fluxes
and the modeling using a point source illumination of the photosphere; the
curves falling within the observed range of \Ka flux provide a constraint on the
loop height during the flare, and are generally consistent with a
loop height less than $\approx$ a few tenths of the stellar radius.
Significant variation of the \Ka flux in these time intervals was also found; this is discussed in \S~\ref{sec:fevary}.

\subsection{XRT+BAT spectral analysis \label{sec:jointfit}}
\begin{figure}[htbp]
\begin{center}
\includegraphics[angle=-90,scale=0.5]{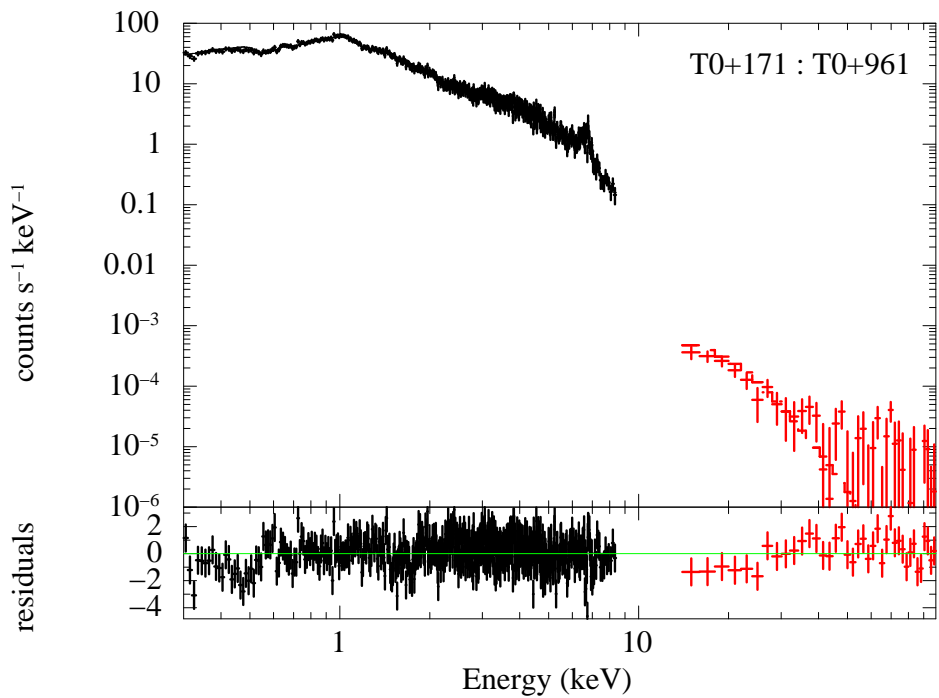}
\caption{Joint spectral fit to XRT and BAT data accumulated over the time interval
T0+171:T0+961.  A three temperature APEC model was used along with a Gaussian line at 6.4 keV.
While a two temperature model was fit to the XRT spectra analyzed on finer time intervals,
the addition of a third component was necessary.  The highest temperature plasma indicated by fitting
the XRT and BAT spectra simultaneously is the same as that from a fit to only the XRT spectrum, and there is 
no evidence for bremsstrahlung emission from a superhot plasma or nonthermal emission in the hard X-ray spectrum.
\label{fig:xrtbat}}
\end{center}
\end{figure}

During the second pointing, both XRT and BAT data were accumulated, and 
a joint spectral fit was done.  For the BAT data this time interval is T0+165.5:T0+961 s,
while for the XRT this time interval is T0+171:T0+961 s.
We added a third temperature APEC component to explore whether there was any evidence
for other temperature plasma.  The addition of a third temperature to the model fit 
reduced residuals below energies of 1 keV, and converged on a temperature less than the original
two temperatures.  Table~\ref{tbl:xrtbat} displays the fit results from 2- and 3-T APEC models to both
XRT and XRT+BAT spectra.  Figure~\ref{fig:xrtbat} displays the 3-T APEC model fit to the XRT and BAT
spectra.
No subsequent improvement in the fit  was gained by adding another spectral model,
such as a higher temperature
bremsstrahlung component or a nonthermal hard X-ray emission component, as was
found for the large II~Peg flare detected by Swift \citep{osten2007}.
A comparison of the fit to only the BAT spectrum reveals a higher temperature (10.0$^{+5.6}_{-2.8}$ keV, 90\% error
range) 
than the larger temperature in the two-temperature fit to the XRT+BAT data (6.66$^{+0.34}_{-0.24}$ keV, 90\%
error range) although the 90\% error ranges of these are only slightly inconsistent.
The high temperature returned from the XRT+BAT spectral fit is the same as that returned from
a fit to only the XRT data.
Thus while the highest temperature measured in the flare during the coverage by the
Swift satellite appears to be $\sim$ 12 keV (or in units of 10$^{6}$K, T$_{6}=$140), 
the systematic effect of only having BAT
data to constrain the temperature may skew this value high.

\subsection{UV and Optical Data}
The UVOT and Liverpool Telescope data together constrain the time-scale of the flare
at optical wavelengths --
the Liverpool Telescope data shows that the flare lasted at least 15 minutes,
as the R filter observations showed a still elevated flux at this time.
At late times in the flare decay there is only a small ($\leq$ 0.5 magnitude) offset above the quiescent
magnitudes, with the $u$, $uvw1$, and $uvm2$ filters indicating 
a transient increase at around T0+31000 s.  The latter is largest in the
$uvm2$ filter, showing an enhancement of 1.2 mag, and the increase is restricted
to filters sensitive to wavelengths longer than about 2200 \AA.

Because the $white$ filter has a response in the ultraviolet, and white-light stellar
flares have most of their enhancement in this wavelength range, we use the upper limit
obtained in this filter to provide a constraint on the flare area.
Previous detailed studies of white light stellar flares have determined that
a black-body function with a temperature of $\sim$ $10^{4}$K often provides a good fit
to the observed broadband flux distribution, with an area covering factor $x$ significantly
less than unity.  The observed UV/optical flux contribution due to such flares can be written as\\
\begin{equation}
F_{\lambda} = \frac{xR_{\star}^{2}}{d^{2}} \pi B_{\lambda} (T_{\rm BB})
\end{equation}
where $F_{\lambda}$ is the measured flux at effective wavelength $\lambda$, 
 $R_{\star}$ the stellar radius, $d$ the distance to the star,
$T_{\rm BB}$ the black-body temperature, and $B_{\rm \lambda}(T_{\rm BB})$ the 
Planck function evaluated at wavelength $\lambda$ and temperature $T_{\rm BB}$.
We used the WebPIMMS\footnote{http://heasarc.gsfc.nasa.gov/Tools/w3pimms.html}
 tool to calculate the integrated flux in the $white$ filter,
with an input black-body temperature of 10$^{4}$K, using the count rate 
of 300,000 counts s$^{-1}$ 
at which the UVOT detector safed.
The flux in the 1600--8000 \AA\ wavelength range from T0+172.9:T0+174.3 s is $\geq$5.5$\times$10$^{-8}$
erg cm$^{-2}$ s$^{-1}$.
We estimate the average
flux over this wavelength range to be $\geq$8.6$\times$10$^{-12}$ erg cm$^{-2}$ s$^{-1}$ \AA$^{-1}$,
and an area covering factor $x$ of $\geq$ 0.03 for the effective wavelength of the $white$
filter 3471 \AA.  For the radius of EV~Lac, R$_{\star}$ is roughly 0.36 R$_{\odot}$ (see discussion in \S3.5), and this works out to an area
of $\gtrsim$2$\times$10$^{19}$ cm$^{2}$.
We also calculated the range in $x$ for black-body temperatures from 8000 K to 20,000 K, finding
that $x$ varies from $\geq$ 0.08 to 0.003 for this range of temperature.
\citet{hawley2003} determined T$_{\rm BB} \approx$ 9000 K
and $x \approx$ 10$^{-4}$ for several flares on the nearby flare star AD~Leo, finding
that each flare in their study had a similar blackbody temperature regardless
of flare energy, peak flux, or total duration.
If we assume that this flare had a similar blackbody temperature, it indicates an area covering
factor $\approx$480 times the well-studied, but smaller amplitude, flares seen on AD~Leo.

We used the WebPIMMS
 tool to calculate also the integrated flux in the $v$ filter at 
T0+156.58:T0+164.78 s
with an input black-body temperature of 10$^{4}$K, and converting the
observed $v$ magnitude of 7.2$\pm$0.2 into a count rate
using the zero-point of 17.89.
The 5000--6000 \AA\ flux from T0+156.58:T0+164.78 s is 5.70$\times$10$^{-6}$
erg cm$^{-2}$ s$^{-1}$, while the 0.3-10 keV X-ray flux near this time 
(T0+171:T0+306 s) is 2.45$\times$10$^{-8}$
erg cm$^{-2}$ s$^{-1}$, leading to an X-ray/optical flux ratio $f_{X}$/$f_{v}$ of 
4$\times$10$^{-3}$.
The quiescent X-ray flux is about 5.8$\times$10$^{-12}$ erg cm$^{-2}$ s$^{-1}$, and the
flare enhancement was 2.9 magnitudes in the $v$ filter, allowing us to compare
this ratio with quiescent values.  It is about 280 times higher than in quiescence.
These values may assist others in trying to make identifications of likely flare
star targets based on optical and X-ray flux ratios.

\subsection{Hydrodynamical modeling of the flare decay \label{sec:hd}}
\citet{reale1997} have described a method to estimate the coronal loop length
from the decay of a stellar flare.  The technique uses temporal information on the rate of decrease
of the flare intensity during the decay phase, as well as time-resolved
spectral analysis to describe the time dependence of the flaring plasma's
temperature and density (which can be estimated from the time dependence of the volume emission
measure, under the assumption that the flare geometry and hence volume does not
change appreciably).  Most flares exhibit an exponential decay, where the decay time 
is determined by the  
balance between cooling from radiation and conduction, as well as 
the presence or absence of sustained heating.
The slope of data points in the temperature-density plane during the flare decay traces the
heating time scale, which can be used to estimate the loop length for the assumption of
a single semi-circular loop with constant cross section.  By performing hydrodynamic modeling of decaying 
flaring loops with a range of dimensions and time-scales, and folding the modeled flare emission 
through the response of a specific instrument, the method can be applied to the
particular bandpass and instrumental response.  This approach has been calibrated 
with resolved solar flares \citep{reale1997} showing good agreement.  It has also been 
applied to a few singular cases of stellar flares with additional length scale constraints,
showing general consistency \citep{favata1999,testa2008}.

This method has been applied to the Swift/XRT bandpass from 0.3-9.5 keV for
a global coronal abundance $Z$ of 0.4 times the solar value.
The thermodynamic loop decay time can be expressed \citep{serio1991} as \\
\begin{equation}
\tau_{th} = \alpha l/\sqrt{T_{\rm max}}
\end{equation}
where $\alpha$=3.7$\times$10$^{-4}$ cm$^{-1}$ s$^{-1}$K$^{1/2}$, $l$ is the loop half-length in cm,
and $T_{\rm max}$ is the flare maximum temperature (K), \\
\begin{equation}
T_{\rm max} = 0.0261 T_{\rm obs}^{1.244}
\end{equation}
and $T_{\rm obs}$ is the maximum best-fit temperature derived from single temperature fitting
of the data. 
The ratio of the observed exponential light curve decay time $\tau_{LC}$ to the thermodynamic
decay time $\tau_{th}$ can be written as a function which depends on the slope $\zeta$ of the decay
in the log(n$_{e}$-$T_{e}$) plane (or equivalently, $\log(\sqrt{VEM}-T)$ plane): \\
\begin{equation}
\tau_{LC}/\tau_{th} = \frac{c_{a}}{\zeta -\zeta_{a}} + q_{a} = F(\zeta) \;\;\; .
\end{equation}
The parameters $q_{a}$, $c_{a}$, and $\zeta_{a}$ are determined for the Swift/XRT instrument
to be $q_{a}$=0.67$\pm$0.33, $c_{a}$=1.81$\pm$0.21, and $\zeta_{a}$=0.1$\pm$0.05. Combining 
the above expression with the one for the thermodynamic loop decay time, a
relationship between flare maximum temperature, light curve exponential
decay time, and slope in the density-temperature plane can be used to estimate the
flare half-length: \\
\begin{equation}
l=\frac{\tau_{LC} \sqrt{T_{\rm max}}}{\alpha F(\zeta)} \;\;,
\end{equation}
valid for $0.4 \leq \zeta\leq 1.9$.  The uncertainty of the loop length is determined by
propagating errors from the $\tau_{LC}$, $\zeta$, $T_{\rm obs}$ values determined from data,
and the uncertainties of $c_{a}$, $\zeta_{a}$, and $q_{a}$ from the numerical calibration
of the method to Swift/XRT data.

\begin{figure}
\begin{center}
\includegraphics[scale=0.9]{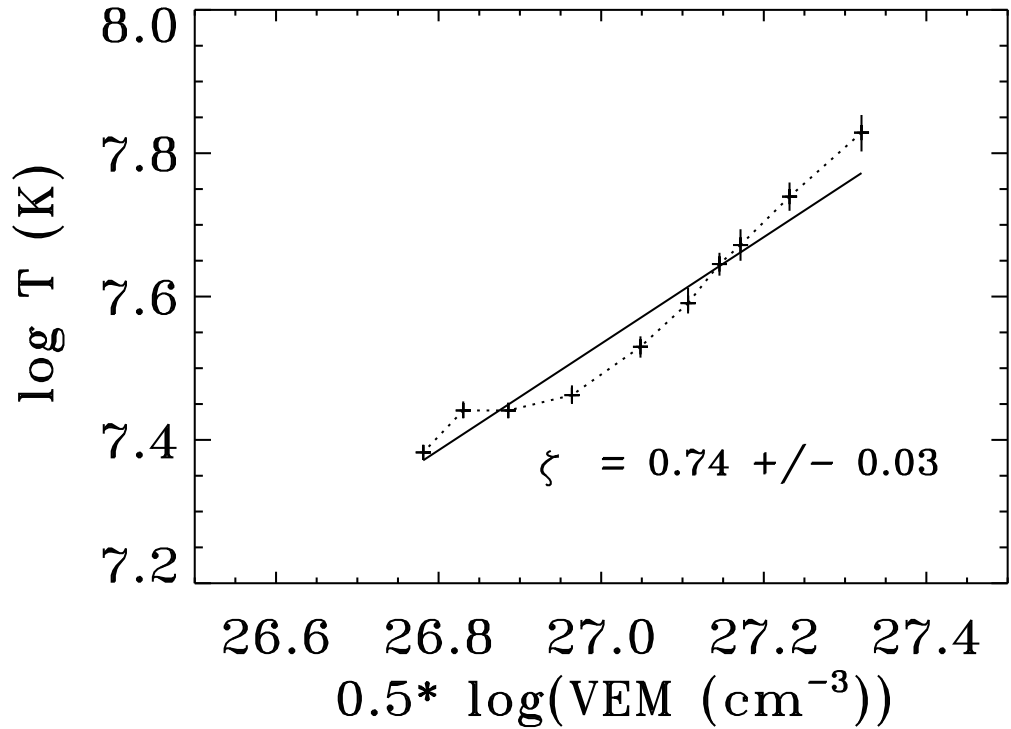}
\caption[]{Variation of temperature versus square root of emission measure for
times in the decay phase of the flare, using XRT spectra in the time range T0+171 s:
T0+2798 s.  Results have been derived from fitting XRT spectra to isothermal
APEC models.
Temporal
evolution proceeds from the top right to the lower left.
\label{fig:ktvem} }
\end{center}
\end{figure}

Section~\ref{sec:xrtspec} discussed the time-resolved spectral analysis of the XRT data.
In order to obtain a density-temperature path, the method requires single values of
flare temperature and emission measure which can be obtained with a $1T$ fitting.
The same fitting is applied to the flaring loop simulations used to calibrate the method.
In this case, the fit provides an effective temperature weighted average on the emission
measure components.
The slope $\zeta$ in the $\log(\sqrt{VEM}-T)$ plane using the isothermal fits
is $\zeta=0.74\pm$0.03 (as shown in Figure~\ref{fig:ktvem}), with a maximum observed temperature $T_{\rm obs}$ of
67$\pm$4 MK.  This corresponds to $T_{\rm max}$ of 142 MK.
Flares in which the time-scale for the decay of plasma heating is gradual have small values of $\zeta$, near 0.4,
whereas flares in which the heating is shut off abruptly have larger values of $\zeta$, near 1.9.
The exponential decay time of the XRT light curve for the first $\approx$2800 s
is $\tau_{LC}=$1014$\pm$5 s. These can be combined using the equations above
to arrive at a loop semi-length estimate $l$ of 9.3$\pm$1.4$\times$10$^{9}$ cm.
The relatively short flare duration and the slope $\zeta$ significantly larger than
the low asymptotic value (0.4) both support a single loop flare rather than a multiple-loop
two-ribbon flare \citep{reale2003}.
Figure~\ref{fig:ktvem} displays a steeper value of $\zeta$ in the initial stages of the
flare decay, closer to a value of 1.1.  Discussion in \S 4.1 showed that the flare decay
had an initially fast decline in the first 1000 s, followed by a slower decline ($\tau_{LC}$
of 647 s followed by $\tau_{LC}$ of 1042 s).  Using these values of $\zeta$ and
$\tau$ to determine the loop semi-length gives consistent results with the 
method above.

Expressing the loop length estimate as a fraction of EV~Lac's stellar radius
requires a precise estimate of the star's radius.  Based on 
mass-luminosity relationship arguments, EV~Lac's mass is likely $\approx$
0.35 M$_{\odot}$ \citep[see discussion in ][]{favata2000}. EV~Lac's spectral type 
of M3.5 places it at the dividing line separating fully convective stars from
those with a thin inner radiative zone.  The exact radius will depend on metallicity
and possibly also on other factors: recent work has shown a systematic discrepancy of
stellar radii of low mass stars determined from eclipsing binaries compared with
low-mass stellar isochrones \citep[][and references therein]{fernandez2009}.
The models of \citet{cb1997} give a stellar radius of 0.36 R$_{\odot}$
for a 0.35 M$_{\odot}$ solar metallicity star older than about 300 MYr.  We add in an uncertainty of 10\% to
this value of stellar radius to account for any systematic effect not included in the 
model determination.  With this, the loop semi-length estimate expressed as a relative
length is $l/R_{\star} =$ 0.37$\pm$0.07. 
If the flaring loop has a semicircular geometry and stands vertically on the stellar surface,
then the maximum loop height $h$ can be expressed as
$h/R_{\star}$=0.24$\pm$0.04.
Since the loop might be inclined on the surface, this value of $h$ should be considered
as an upper limit.

\subsection{Fe K$\alpha$ 6.4 keV and Its Variations\label{sec:fevary}}
To date there have been only a handful of detections of Fe K$\alpha$ 6.4 keV emission
in stars without disks \citep{osten2007,testa2008}, and these have occurred in evolved
stars.  The best studied case, due to our ability to spatially resolve flaring
regions, is the Sun \citep{bai1979},
where the hot continuum X-ray radiation from a flaring coronal loop-top source
 illuminates the photosphere and 
causes fluorescence of photospheric iron.  
In addition, there have been more detections of Fe K$\alpha$ emission in young stars, where the 
possible presence of a disk complicates the geometry, with reflection of the stellar
X-ray continuum off the surrounding disk likely being a dominant contributor \citep{tsujimoto2005}.
EV~Lac does not appear to have a disk \citep{lestrade2006}, and thus the detection of the fluorescence line
is the first time this has been seen in a main sequence star other than the Sun.

The \Ka feature is detected most significantly in the early stages of the flare decay,
and so we investigated the time-dependent characteristics and formation of the line.
Due to the need to mitigate the pile-up effects discussed in \S 2.3, there is not enough
signal on shorter time intervals than the time bins we have used in Table~\ref{tbl:fek}.
Extracting spectra over larger time intervals does not improve the significance of 
\Ka equivalent width much,
indicating that it is likely fluctuating on shorter time-scales compared to the size of the larger bins.
In the first five time intervals 
the photoionizing hot continuum, defined as $N_{\rm 7.11}/4\pi d^{2}$, declines by a factor
of 6, while the plasma cools from 8.4 keV to 4.8 keV. This results in a general decline
in the strength of the \Ka feature.  The \Ka line fluxes through the decay phase of the flare
are generally consistent with loop heights on the order of a few tenths of a stellar radius.
The hydrodynamic modelling of \S~\ref{sec:hd} returned an estimate of the 
height of the flaring loop of $h/R_{\star}=$0.24$\pm$0.04.  
Thus the two methods are in agreement on the coronal loop height
within the uncertainties of factors of a few.  

The time variation of the \Ka line flux is generally consistent with 
fluorescent emission from compact loops and the cooling
of the flaring plasma 
occurring during the decay of the flare.
However, the initial stages of the flare decay show some 
marked variations on short time scales.
For 135 s (T0+171:T0+306 s)
the \Ka flux is detected at $>$99\% confidence, indicating a compact
loop of height $<$0.1 R$_{\star}$, and then in the next 202 s (T0+306:T0+508 s)
the line is undetected, with a lower limit of zero flux.  In the third time segment, spanning
163 s (T0+508:T0+671 s), the line flux is again present but at a level 
greater than what is predicted based on fluorescence from the derived loop height --- in fact, the observed \Ka flux
is not consistent with any set of fluorescent models, even for those that go to zero height.  
In this third time interval, the most compact loops (with $h$ asymptoting to zero height)
would have a \Ka line fluorescent flux which is $\approx$8$\times$10$^{-3}$ photons
cm$^{-2}$ s$^{-1}$. 
The strength of the \Ka flux is about three times larger than that, with an total flux of about
2.64$\times$10$^{-2}$ photons cm$^{-2}$ s$^{-1}$.   The excess emission is conservatively in the 
range 0.0089 $<$ F$_{6.4 keV}$ $<$ 0.028 photons cm$^{-2}$ s$^{-1}$ (using the 3$\sigma$ error bars
in Table~\ref{tbl:fek}).
The time-scale for variation of this
flux is rather short, being present for only 163 s.  
The duration is short enough that changes in astrocentric angle (such as may happen during 
a large change in phase of stellar rotation) can be excluded.   The fluorescent efficiency $\Gamma$ depends on 
the fluorescing spectrum, flare height, and relative iron abundance.  \citet{drake2008}
show that $\Gamma$ varies very slowly with temperature, decreasing by a factor of 2 over
a factor of 30 in temperature. The temperatures at which the K$\alpha$ feature is detected
are 6--8 keV (70--90 MK) and the integrated photon density spectrum decreases by a factor
of 3 during the first three time intervals.
The relative fluorescent efficiency varies with abundance by weaker than a proportional relation \citep{drake2008},
and there is no evidence for abundance variations taking place during the decay of this flare.

One possibility that has been used to explain Fe \Ka emission during the impulsive phase of a few solar flares
is the association of the \Ka line flux variations with hard X-ray bursts \citep{tanaka1984,emslie1986,zarro1992}. 
\citet{tanaka1984} attributed the intense \Ka emission in the early phases of the solar flare to fluorescence
from the power-law hard X-ray bremsstrahlung emission, noting that further \Ka emission after the decay
of the hard X-ray flux was consistent with fluorescence from a thermal source.
\citet{emslie1986} determined in the solar flare they considered that the excess \Ka emission above that produced by thermal photoionization
was consistent with collisional ionization from accelerated electrons, and 
\citet{zarro1992} noted during the impulsive phase of the solar flare they analyzed 
that a thermal photoionization interpretation of the observed \Ka line flux could not account for all of
the observed \Ka line flux, requiring a component from collisional ionization.
In contrast to the situation presented in these solar flares, in the present case we are
detecting \Ka line emission during the decay phase of the stellar flare, and the variability
appears during this decay phase.
As noted in \S~\ref{sec:jointfit}, the analysis of the combined XRT and BAT spectra during the flare
decay over the time interval T0+171:T0+961 s showed no evidence for an additional spectral component beyond
the two temperature APEC model and the \Ka line emission at 6.4 keV.  
The BAT data did not have enough counts to extract spectra over shorter time intervals, and so we
use this spectra to constrain the action of nonthermal electrons over the larger time interval.

\begin{figure}
\begin{center}
\includegraphics[scale=0.8]{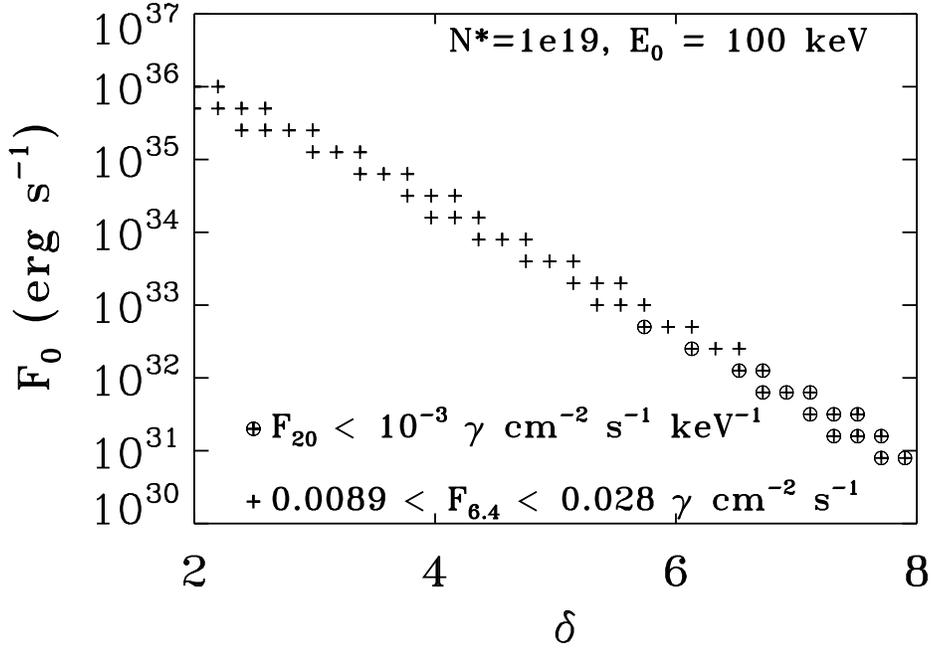}
\caption[]{ A grid of parameter values for thick target bremsstrahlung
emission was explored to determine if this was a viable mechanism to explain the
excess \Ka flux during the time interval T0+171 s:T0+961 s. 
Values of the total power in the electron
beam F$_{0}$ are plotted against the power-law index of the electron
distribution function $\delta$, for a cutoff energy E$_{0}=$ 100 keV,
and column density in the \Ka-emitting layer
$N^{*}=$10$^{19}$.
 Plus symbols indicate values compatible with the
excess \Ka flux measured.  Plus symbols over-plotted with open circles are grid points
compatible with both the excess \Ka flux and a 20 keV photon flux less than the amount 10$^{-3}$
photons cm$^{-2}$ s$^{-1}$ observed in the BAT spectrum.  See \S 3.6 for more details.
\label{fig:kacoll}
}
\end{center}
\end{figure}

Following the discussion in \citet{emslie1986}, we can determine the amount of Fe \Ka emission produced by
collisional ionization.  The observed \Ka flux depends on the characteristics of the 
distribution of accelerated electrons, namely the power-law index $\delta$, the total power
in the beam $F_{0}$, and the low-energy cutoff $E_{0}$. Additionally, the \Ka flux depends on the
column density in the \Ka-emitting layer ($N^{*}$), as some fraction of accelerated electrons will
lose their energy via collisions in dense regions. We computed the expected \Ka flux on
a grid on $\delta$, $F_{0}$, and $N^{*}$, and cutoff values $E_{0}=$20 keV and $E_{0}=$100 keV,
and found ranges of parameter space where the \Ka flux due to collisional ionization was at the level
of the \Ka flux observed in the third time interval --- the
measured \Ka flux and errors
minus the maximum amount expected from a fluorescence process
(0.0089 $<$ F$_{6.4 keV}$ (photons cm$^{-2}$
s$^{-1}) <$ 0.028).
We then used these parameters to estimate the amount of nonthermal thick-target
bremsstrahlung flux.  
Since the observed flux level at 20 keV during the time interval T0+171:T0+961 s is
$\approx$10$^{-3}$ photons cm$^{-2}$ s$^{-1}$ keV$^{-1}$ and there is no evidence
for nonthermal contribution to this flux, we used this value 
to determine plausible ranges of parameters which could both produce enough \Ka
emission to provide the required extra \Ka flux and not be detected in the BAT spectrum.
For column density in the \Ka-emitting layer
of 10$^{19}$ cm$^{-2}$ or less, low energy cutoff E$_{0}$ of 100 keV, values of total power in the beam between 10$^{30}$ and 10$^{33}$
erg s$^{-1}$, and $\delta$ between 6 and 8, the photon flux at 20 keV is $<$ 10$^{-3}$
photons cm$^{-2}$ s$^{-1}$ keV$^{-1}$.
Figure~\ref{fig:kacoll} displays the
range of parameter space, with pluses indicating where the expected flux was within the measurement
range for the excess \Ka emission line flux, and the pluses with circles around them indicate parameter values consistent 
with a 20 keV thick target bremsstrahlung flux of $<$ 10$^{-3}$ photons cm$^{-2}$ s$^{-1}$.
Thus it is plausible that the excess emission
in the \Ka line could be produced by collisional ionization, and yet not be detected as an additional
spectral component in the BAT spectrum.
There are examples of solar flares which have high values of the low energy cutoff $E_{0}$
\citep{warmuth2009} and high values of $\delta$ \citep{sui2004}.


The optical area constraints from the white light flare discussed in \S 3.4 lead to an area A $\gtrsim$ 2$\times$
10$^{19}$ cm${^2}$ .
This is the area of the photosphere involved in the flare, and if we interpret this as the foot-point area, then 
we can combine this measurement with the above determined ranges of total electron beam power, to arrive at an
estimate of the nonthermal beam flux.  For a beam power between 10$^{30}$ and 10$^{33}$
erg s$^{-1}$, this leads to a nonthermal beam flux of 10$^{11}$--10$^{14}$ erg cm$^{-2}$ s$^{-1}$.
The values typically found in solar flares are 10$^{10}$ erg cm$^{-2}$ s$^{-1}$, with 10$^{11}$ being 
found in some large flares.
\citet{allred2005} discuss radiative hydrodynamic models of solar flares, and a later paper
\citep{allred2006} extended the results to M dwarf stellar flares; they found that for the same
beam fluxes, the time-scales in both the solar atmosphere and M dwarf atmosphere were shorter for
larger beam fluxes.  For a beam flux of 10$^{11}$erg cm$^{-2}$ s$^{-1}$, the dynamics of the 
impulsive phase of the flare lasts $<$ 15 seconds in both cases.  
These short time-scales are consistent with the short time-scales observed in the flare on EV~Lac.

%
\section{Discussion}

\subsection{Flare Energy Estimates}
Because of the heterogeneous wavelength and temporal coverage during the flare decay, we
made flare energy estimates a few different ways.
The Swift observations cover the initial $\sim$3000 s of the flare decay 
and  additional observations extending up to 4$\times$10$^{4}$ s after the trigger.
Table~\ref{tbl:energy} lists the determination of radiated energy at various
times and wavelength ranges during the flare. 
We estimated the energy radiated in the 0.3--10 keV range during the time intervals in which only BAT spectra
were obtained by extrapolating the best fit to the 15-100 keV X-ray spectrum into this energy range.
We also examined the temporal evolution of the 0.3--10 keV flux as shown in Figure~\ref{fig:specpar}.
Fitting this with a double exponential function with a break around 1000 s after
the trigger, we find an initial exponential decay $\tau_{1}$ of 647 s and second decay
$\tau_{2}$ of $\sim$ 1042 s. We then extrapolate to the point where the 
flux would equal the measured quiescent X-ray flux, $\approx$5.8$\times$10$^{-12}$
erg cm$^{-2}$ s$^{-1}$, which is $\sim$ 8300 s after T0.  
In the 0.3-10 keV band we estimate the radiated
energy to be $\approx$7.3$\times$10$^{34}$ ergs from T0:T0+8300 s.
The flux measured during the time span 6160--19792 s after T0
is 10 times higher than the quiescent flux, even though extrapolation of the
flux due to an exponential flare decay would indicate a return to quiescent conditions
at around T0+8300 s. 
This suggests that the flare likely
was more complex, with a re-brightening during this gap in time, and
a continuation beyond the Swift monitoring time.
Since this was apparently a long-duration flare, 
our estimates of the radiated flare energy
are actually lower limits to the total energy radiated in X-rays. The flare radiated
energies are at the high end of what has been observed for flares on other M dwarfs.
Note that the exponential time-scales derived from BAT and Konus in equivalent time ranges are shorter:
from T0:$\sim$T0+300 s $\tau$ is 175$^{+17}_{-14}$ s for BAT (14--30.1 keV)
and 187$\pm$7 s for Konus (18-70 keV).

We note that in the first spectrum of the flare, from (T0-39:T0 s),
$\sim$62\% of the stellar bolometric flux was recorded at hard X-ray energies (14--100 keV).  
By extrapolating
the temperature and emission measure determined from that spectral analysis to smaller X-ray energies,
we estimate that in the XRT band (0.3--10 keV) the flux may have been as high as 4.1$\times$10$^{-8}$
erg cm$^{-2}$ s$^{-1}$, or $\approx$ 2.4 times the bolometric flux from the star.
This is about 7000 times larger than the quiescent flux estimated by \citet{osten2005}.
Thus from 0.3--100 keV the peak luminosity was 1.6$\times$10$^{32}$ erg s$^{-1}$, or
a value L$_{X}$/L$_{\rm bol}$ of $\sim$3.1, from T0-39:T0 s.
The  upper right panel of Figure~\ref{fig:specpar} displays the bolometric
flux, along with the temporal evolution of the flux in the 0.3-10 and 15-100 keV energy
ranges; for the first $\approx$ 500 s after the trigger the X-ray flux exceeded
the bolometric flux.

At first glance it may seem problematic to have a flare whose luminosity exceeds the normal
stellar bolometric luminosity.  The free energy which powers the flare  
is built up over long time-scales in a more or less continuous manner.  However, that energy
can be released suddenly through explosive magnetic reconnection.  
The largest energy flares
which can be produced relate to the largest active region size which can exist
\citep{aschwanden2007}.
This flare was an extreme example of instantaneous magnetic energy release,
but the radiated energy in the X-ray band
is typical of longer duration flares. 

\subsection{Comparison to Other Superflares}
While there have been
stellar flares detected above 10 keV in the past (mostly with the BeppoSAX satellite), these Swift/BAT
flares \citep[the one on EV~Lac reported here and the superflare on II Peg reported by][]{osten2007}
are different in that they were observed using autonomous triggers based on their hard X-ray flux
rather than in pointed observations which serendipitously detected large stellar flares
\footnote{Since this flare, there have been two additional autonomously triggered stellar flares
observed with Swift: http://gcn.gsfc.nasa.gov/gcn3/8371.gcn3 and http://gcn.gsfc.nasa.gov/gcn3/8378.gcn3;
these are the subject of future papers.}.
The two triggered flares differ in several important 
respects:  the maximum plasma temperature achieved was very large 
in the early stages of the flare on II~Peg (peaking at $\sim$180 MK), and the temperatures
measured with both XRT and BAT spectra at about 3 minutes past the triggers indicate hotter
temperatures for the case of II~Peg versus EV~Lac (120 MK and 100 MK, respectively).  
The II~Peg flare had a longer rise time (1000--2000 s) compared with $<$120 s for EV~Lac, and
while neither flare was observed in its entirety, the decay of the flare on II~Peg had a slower decline
than the flare on EV~Lac.
Some of these differences are likely due to the different stellar environments: 
II~Peg is a tidally locked binary system consisting of a K subgiant and M dwarf companion with a
likely age of a few Gyr,
while EV~Lac is a younger single M dwarf flare star, with a much higher gravity.

At its peak, the EV Lac flare is $\sim$2.5 times II Peg's maximum 14-40 keV hard X-ray count rate, and the
XRT flux is $\sim$ factor of 4 larger.
Due to the different distances (42 pc for II~Peg vs. 5.06 pc for EV~Lac), the X-ray luminosity for II~Peg during the peak 
of that flare 
was larger, at 1.3$\times$10$^{33}$
erg s$^{-1}$ (0.8--10 keV), compared with 6.5$\times$10$^{31}$ erg s$^{-1}$ for EV~Lac in the
same energy band. Expressed as a fraction of the star's bolometric luminosity, though, 
the EV~Lac flare is larger, with $L_{\rm 0.8-10 keV}$/L$_{\rm bol}$ $\approx$ 1.2, compared to 
L$_{\rm 0.8-10 keV}$/L$_{\rm bol}$ = 0.24 for II~Peg.
As discussed in \S 4.1, up to several times EV~Lac's normal bolometric luminosity  may have been
radiated at X-ray energies when estimating the soft X-ray flux in the early stages of the flare.
Another crucial difference is the existence of additional emission in the hard X-ray band in II Peg,
which \citet{osten2007} interpreted as nonthermal emission; in the EV~Lac flare there is
no such evidence.
Both flares revealed evidence for Fe K$\alpha$ 6.4 keV emission, but the EV~Lac flare
sports a larger maximum equivalent width and photon flux.
While \citet{osten2007} described the formation of the \Ka line flux seen in II~Peg 
as originating entirely from collisional ionization, \citet{drake2008} showed
that a fluorescence interpretation is more likely.  The \Ka line flux shows temporal
variations in both flares, and here we interpret the excess \Ka emission 
as suggestive of the action of nonthermal electrons.

Besides the two flares discussed above, two other flares have detections in the 10--50 keV band
made with BeppoSAX --- 
a flare on the RS~CVn binary UX~Ari \citep{franciosini2001}, and a flare on Algol
\citep{favata1999}. The common characteristics of these flares were a super-hot temperature
(peak temperatures in excess of 100 MK), large X-ray luminosity (generally $\sim$10$^{32}$
erg s$^{-1}$, in different energy bands), and long duration, as measured by the time to
decay from peak count rate to half that value ($\sim$ 8 and 12 hours for the Algol and UX Ari flare,
respectively). The flare on UX~Ari showed no abundance variations, while the one on Algol did show
factor of $\sim$3 abundance enhancements in the early stages of the flare.
Observations of stellar flares with the Ginga satellite have detected X-ray emission out
to 20 keV in a few large flares. \citet{pan1997} found peak temperatures 
near 100 MK and and emission measures near 10$^{54}$ cm$^{-3}$ from a large flare
identified with the M dwarf star EQ1839.6+8002, and flare peak temperatures from 60--80 MK
were determined for flares on the active binaries Algol, UX~Ari, and II Peg 
\citep{sternginga,tsuruginga,doyleginga}.

ATEL \# 1689 \footnote{http://www.astronomerstelegram.org/?read=1689}
and \# 1718 \footnote{http://www.astronomerstelegram.org/?read=1718}
describe an X-ray transient detected by XMM-Newton which may be an M star.  With a distance of 400 pc,
its peak X-ray luminosity in the 0.2--10 keV range was 5.5$\times$10$^{32}$ erg s$^{-1}$.
The authors determine the spectral type of the star to be M2/3, and thus
this X-ray flare, if associated with this star,
 would likely have exceeded its bolometric luminosity ($\sim$10$^{32}$ erg s$^{-1}$).

\subsection{Comparison to Other Flares on EV~Lac}
Prior to the large flare discussed in this paper, Swift/BAT detected lower amplitude 
hard X-ray transient emission from EV~Lac on three occasions which were not strong enough to
trigger an autonomous slew.  ATEL \#1499 \footnote{http://www.astronomerstelegram.org/?read=1499}
describes these.  The peak signal-to-noise ratio in the BAT data of the autonomously
triggered stellar flare was 19.47,
while two possible events on 6 June 2006 and 20 May 2007 had SNR of 3.5 and 3.1 sigma, respectively.
An event on 22 April 2008 was a 4.0 sigma detection over 64 seconds, and is a likely flare detection.

EV~Lac earlier underwent a large flare which was detected by the ASCA satellite, 
discussed by \citet{favata2000}.
The maximum plasma temperature diagnosed was 6.3 keV ($\approx$ 70 MK) with a peak X-ray
luminosity (0.1--10 keV) of $\sim$10$^{31}$ erg s$^{-1}$ (25\% of L$_{\rm bol}$), and radiated energy
at X-ray wavelengths (0.1--10 keV) of 1.5$\times$10$^{34}$ erg ($\approx$ 300 times the 
quiescent level).
Like this flare, the one observed by ASCA was relatively short, with an exponential
decay of 1800s, and came from a relatively compact flaring loop ($l\lesssim$0.5R$_{\star}$).
Visual inspection of the Konus-Wind data for the time of the flare caught by the ASCA satellite 
shows no significant emission, suggesting the 
flare detected by Swift was harder and more luminous in the hard X-ray range.
Another key difference between the flares was the presence of significant abundance
variations in the ASCA flare (factor of $\approx$6 between maximum and minimum $Z$).

A moderate flare on EV~Lac seen with the Suzaku X-ray telescope 
lasting 1500s was discussed by \citet{evlacsuzaku}.  In comparison with the 
spectrum extracted outside of flaring intervals, the abundance during the flare appears to have
increased by factors of several.  This behavior is similar to what was observed in the large
ASCA flare.
\citet{schmitt1994} discussed a flare seen on EV~Lac in the ROSAT All-Sky Survey which lasted 24 hours,
with a peak temperature near 25 MK.  Numerous small flares have also been observed with Chandra ---
\citet{osten2005} described nine small flares, with highest temperatures near 30 MK, and
\citet{huenemoerder2009} discussed another Chandra observation exhibiting numerous
low-level flares.





\subsection{Coronal Length Scales}
The determination of coronal length scales provides an important constraint on the
energetics and dynamics of coronal processes.
The use of hydrodynamic modelling to describe the variation of plasma temperature and volume
emission measure through the flare decay phase allows an estimate of the coronal loop
length to be made, assuming that a single flaring loop dominates the emission.
Likewise, the detection of the Fe \Ka line allows a constraint to be made on the height above
the surface of a coronal loop illuminating the cold iron which fluoresces to form the \Ka line.
While the hydrodynamic modelling explicitly assumes that a single loop is involved in the flare, 
the \Ka emission line strength depends mainly on the height.  The fact that the two are consistent to within
a factor of a few suggests that the former assumption is roughly consistent with the data.
%

Loop lengths derived from an analysis of the time variation of flare light curves
in conjunction with variation of temperature and emission measure using hydrodynamical modelling
of decaying flaring loops indicates relatively short loops for 
two large flares observed on EV~Lac: \citet{favata2000}
derive a loop semi-length $l$ of $\sim$0.5R$_{\star}$ during a large flare
seen with the ASCA satellite, while in the present case a 
comparable loop length is derived, $l \sim$ 0.37R$_{\star}$.

Another method to estimate length scales involves resonance scattering in emission lines,
where the expected flux of density-insensitive resonance lines in emission 
departs from that expected under optically thin conditions.
Then the scattering length can be deduced if the electron density in the medium is known.
\citet{testa2007} determined from time-averaged coronal spectra obtained with
Chandra's HETGS of EV~Lac
that the ratio of Lyman-$\alpha$ to Lyman-$\beta$ lines
of \ion{O}{7} was more than 3$\sigma$ below the theoretical value assuming optically thin emission.
Using an escape probability formalism they estimated a path length of 1.6$\times$10$^{9}$ cm
(0.06R$_{\star}$)
for EV~Lac's O~VII-emitting material.
Such techniques rely upon the measurement of individual lines, which is not possible with the 
XRT on Swift.  These length scales are smaller than those determined for the flaring loop in this case,
but are within an order of magnitude.

\begin{figure}[htbp]
\includegraphics{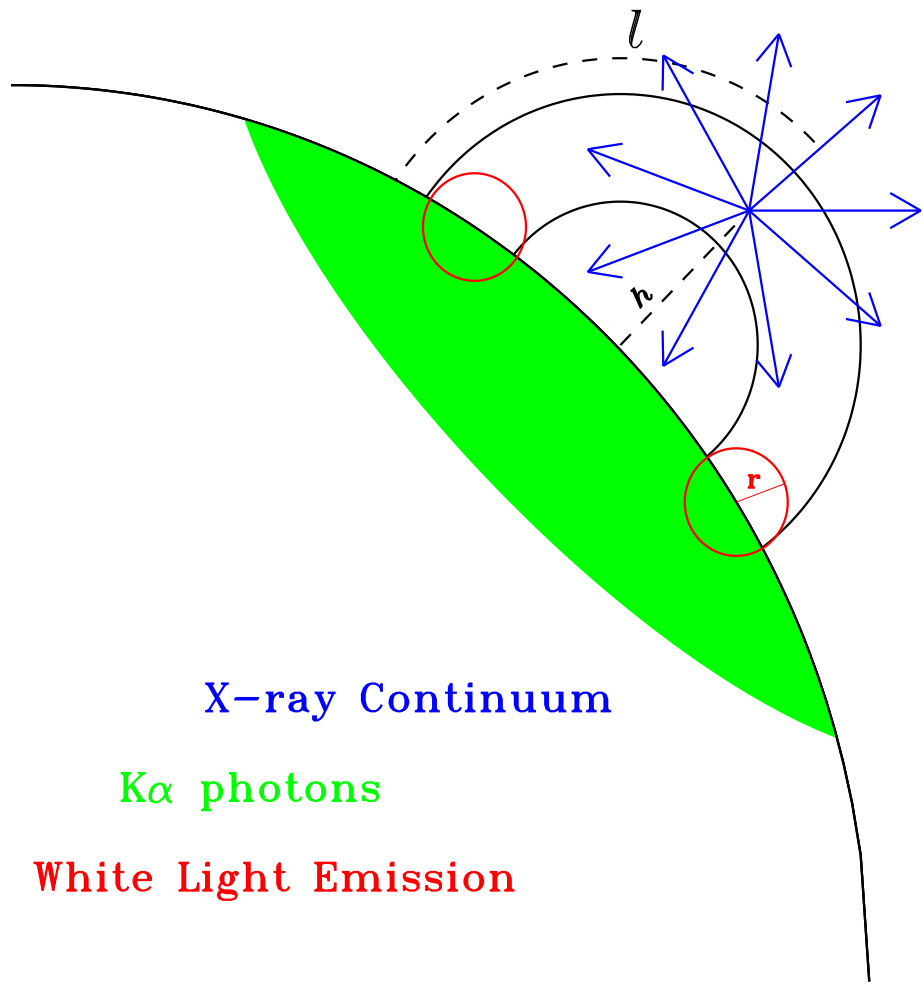}
\caption{Schematic showing the geometry between the flare photons, emitted as
high energy X-ray continuum, the region where the photospheric \Ka photons are emitted,
and the probable location of the white light flare.  Different length scales
used in the text are also indicated. The coronal loop is shown scaled to the stellar radius in 
approximate ratio to the values indicated from our analysis, and the foot-point area is a lower limit based on the white light flare emission.
\label{fig:kalpha}}
\end{figure}

If we use the flare area determined from the white light flare and assume that this is the 
area of the foot-point of a single coronal loop, then we can estimate the radius of the coronal loop at the
base under the assumption of simple cylindrical loop.  
From \S 3.4, A $\gtrsim$2$\times$10$^{19}$ cm$^{-2}$, so dividing by two for a contribution from each foot-point, we arrive at
a foot-point radius of $r\approx$10$^{9}$ cm.  The loop length estimates from \S 3.5 give a semi-length of
$l=$9.3$\times$10$^{9}$ cm, and combining the two gives an aspect ratio $\alpha = r/(2l)$ of 0.1.
This is the value commonly assumed in solar flares \citep{golub1980}.
Figure~\ref{fig:kalpha} illustrates the relationship between the X-ray photons, coronal loop,
white-light flare emission, and region of \Ka emission.
The plasma volume emission measure (VEM) contains spatial information through the total volume, which can be used to constrain 
loop length given an electron density, or used to estimate electron density if a loop length is assumed.
With $\alpha$ set equal to 0.1, the volume can be written as \\
\begin{equation}
V= 8 \pi \alpha^{2} l^{3}
\end{equation}
and the VEM as \\
\begin{equation}
VEM = n_{e}^{2} V = 8 n_{e}^{2} \pi \alpha^{2} l^{3} \;\;\; .
\end{equation}
The peak 0.3--10 keV VEM during this flare was $\approx$ 10$^{54}$ cm$^{-3}$.  If we use the 
loop length derived from the hydrodynamic modelling to estimate an electron density,
and assume that there was only one flaring loop, then 
 an upper limit on n$_{e}$ is around 6$\times$10$^{12}$ cm, which is quite high.
It is an upper limit on n$_{e}$ because the loop length might be more than 9.3$\times$10$^{9}$ cm 
and inclined on the surface, so the loop volume might be larger, and hence the density smaller.
We have no constraint from X-rays on the loop number or cross-section area.  
If the number of loops was larger than one, this would not change the observed
plasma evolution.

\subsection{Astrobiological Implications }
Due to their lower luminosity, M dwarfs have a habitable zone which is 5--10 times
closer than for a solar-like star.  This decreased distance renders an exoplanet
more susceptible to influences by the parent star.
It has already been recognized that flares from dMe stars affect the atmospheres
of any terrestrial exoplanets through frequent stochastic ionizing radiation
\citep{smith2004}.
We can quantify the effect that the flare being discussed in this paper would have on an exoplanet in the 
EV Lac's habitable zone
by comparing it to solar flares using the solar X-ray flare classification.
Using the spectral model during the trigger observation (T0-39:T0 s), we
calculate the flux in the 1--8 \AA\ (1.54--12.4 keV) range, and scale from 
a distance of 5.06 pc to 0.1 AU.  The solar flare classification is $Xn = n \times 10^{-4}$ W m$^{-2}$,
and for this flare the scaled flux is 3600 W m$^{-2}$, or an X3.6$\times$10$^{7}$.  For comparison,
the largest solar flare is about X28.

\citet{smith2004} also determined that roughly 4\% of the ionizing X-ray radiation can penetrate
a terrestrial atmosphere, through redistribution into biologically damaging UV radiation.  In the
flare considered here on EV~Lac, the X-ray emission was approximately 7000 times the quiescent level
during the peak of the flare, suggesting that the UV radiation which a putative
terrestrial planet around EV~Lac would have experienced was increased by about a factor of 280 
due only to the X-rays.  Since the X-ray flare was accompanied by white light flare, which was a factor of
76 brighter in the UV than the quiescent UV emission, the total increase in the incident radiation on a
terrestrial planet would have been $\sim$350 times larger.  




The UV flux from a parent star can affect the evolution of a terrestrial planetary atmsophere,
and can be both damaging and necessary for biogenic processes on a terrestrial planet.  \citet{buccino2007} 
defined a UV habitable zone which balances these two competing pressures, and found that for inactive M stars
the  UV habitable zone was closer to the parent star than the traditional liquid water habitable zone.
During moderate flares, however, the temporary increase in UV radiation made the two habitable zones overlap,
suggesting that there may be a beneficial role for frequent, moderate sized flares in supporting 
habitability on exoplanets around M stars.

The computation of habitable zone distance is partly determined by the star's bolometric
flux.  The extreme flare here represents a case of a transient increase in the
star's bolometric flux, with an energy distribution significantly blue-ward
of the quiescent spectral energy distribution.  Although this magnitude of flare is likely
rare, it likely represents the ``tip of the iceberg'' of a whole ensemble of
more frequent but less energetic flares whose integrated emission may
change the time-averaged bolometric luminosity over the value estimated from the luminosity
in the visual band using the standard bolometric correction.
This is an additional factor which 
may influence habitable zone distances for M stars.
Future attempts to characterize the habitability of exoplanets around M stars should take into account
the influence of such extreme events.

\section{Conclusion}
The flare studied in this paper is unique in two main aspects: the level of X-ray radiation
marks it as one of the most extreme flares yet observed in terms of the enhancement 
of $\sim$7000 compared to the usual emission levels, and it is one of only a handful of stellar
flares from normal stars without disks to exhibit the Fe \Ka line, with a maximum 
equivalent width of $>$200 eV.

It is somewhat ironic that although early explanations for the source of GRBs considered
gamma ray emission from flare stars \citep{li1995}, they were quickly swept aside due to 
the extreme nature of the events needed and the apparent discontinuity
with the population of observed stellar flares.  Now that gamma-ray burst detectors have become 
more sensitive, we are indeed finding a small population of very energetic stellar flares which
produce transient hard X-ray flux at levels consistent with those seen from extragalactic GRBs.
These highly luminous events do appear to share characteristics with lower amplitude flares, and
are likely an extension of the same underlying energy release processes.  Thus, the fact that
these flares exhibit
Fe \Ka line emission and evidence (both direct and indirect) for the existence of nonthermal hard
X-ray emission does not imply that there is something special about these flares; 
stated in the solar physics literature  as the ``big flare syndrome'' \citep{bfsref},
the larger the energy release, the more likely one is to see different flare energy manifestations.
Perhaps, the most interesting finding is the detection
of Fe \Ka line emission, as its presence in the X-ray spectra of normal stars without
disks gives an independent constraint on the height of flaring coronal loops,
and provides a consistency check with results from flare hydrodynamic models.  As previous
attempts to characterize the length of flaring loops using different theoretical methods has
led to different values \citep{favata1999}, the use of independent methods is an extremely useful
`sanity check' on these models.

The detection of Fe \Ka line emission in flares from normal stars is even more interesting
in light of its utility in understanding the formation of this
line 
in young stellar objects, where the possible presence of a disk complicates the interpretation.
In young stellar objects, the Fe \Ka line is somewhat more commonly detected, but
due to the often unknown nature of disk geometry around these stars, the interpretation
of the strength of the emission is more ambiguous.  Indeed, \citet{giardino2007}
determined that the observed Fe \Ka line strength in one young stellar object exceeded that 
allowed either by fluorescence or disk emission, and suggested the presence of nonthermal electrons
in the star-disk interaction.  The advantage of the Swift-triggered stellar flares is not
only their high intensity which allows for the detection of this line, but the ability to
study the line flux variability at high time resolution, along with simultaneous constraints
on the hard X-ray flux level.

The size of the flare, in terms of its peak X-ray luminosity exceeding the non-flaring stellar
bolometric luminosity, is impressive.
The existence of such flares can provide important constraints on the time-scales for energy
storage and release in a stellar context.
The multi-wavelength observations which comprise this flare study have enabled the determination
of both the rise time of the flare (Konus/Wind and Swift/BAT constraints) as well as
the photospheric contribution to the flare at early times (Swift/UVOT $v$ and $white$ filter
constraints).  While more extreme stellar white-light flares have been observed from other
M dwarf stars, the combination of the UV, optical, soft and hard X-ray measurements
were key for determining the possible contribution of nonthermal electron beams to 
the Fe \Ka line flux, and confirming the self-consistency of this explanation by estimation 
of nonthermal beam fluxes.  
While we find only indirect evidence for the role of nonthermal electrons in affecting
the X-ray spectrum during stellar flares, this work bolsters conclusions from other
authors that additional processes affect stellar X-ray spectra beyond
the usual contributions of optically thin thermal emission in collisional ionization equilibrium.
The X-ray to optical flux ratios seen in the early stages
of the flare will hopefully be useful in constraining the contribution of flaring
M stars to the population of unidentified transient
hard X-ray sources.

The frequency of occurrence of flares of this magnitude is unknown,
but they may be an important factor in 
determining the habitability 
of planets around M dwarfs, as even a single occurrence of such a large
radiative release can have disastrous consequences for terrestrial-like atmospheres
on exoplanets orbiting M stars.  As progress is made on identifying nearby stars hosting
exoplanets, and because of the renewed emphasis on M dwarfs as host stars, such 
studies must be done in parallel with those attempting to understand the detailed 
nature and distribution as a function of luminosity of the stellar flares themselves.

\acknowledgements
This work made use of data supplied by the UK Swift Science Data Centre at the University of Leicester.
The Konus-Wind experiment is supported by a Russian Space Agency contract and RFBR grant 09-02-00166a.
We thank the Swift team for the ToO on Swift $\sim$ one year after the flare to determine
the quiescent level of EV~Lac in several UVOT filters.

\facility{Swift,Liverpool:2m}
\object{EV Lac}


\clearpage

\begin{deluxetable}{llll}
\tablewidth{0pt}
\tablenum{1}
\tablecolumns{4}
\tablecaption{UV and Optical Magnitudes of EV Lac\label{tbl:mag}}
\tablehead{ \colhead{Filter} & \colhead{Telescope} & \colhead{Time Interval\tablenotemark{a}} & \colhead{Value} \\
\colhead{} & \colhead{} & \colhead{(s)} & \colhead{} }
\startdata
$v$ & Swift/UVOT & T0+156.58:T0+164.78  & 7.2$\pm$0.2 \\
    & 	         & Q (UVOT Tool) & 10.1\\
$white$ & Swift/UVOT & T0+172.9:T0+174.3  & $<$ 7 \\
    & 	         & Q (UVOT Tool) & 11.7\\
$u$ & Swift/UVOT & T0+17670.5:T0+17970.3  & 12.96$\pm$0.01\\
  &            & T0+17974.0:T0+18273.8  & 12.97$\pm$0.01 \\
  &            & T0+18277.6:T0+18577.4  & 12.95$\pm$ 0.01 \\
  &            & T0+24289.8:T0+24589.6  & 12.89$\pm$0.01 \\
  &            & T0+24593.5:T0+24893.3  & 12.93$\pm$0.01 \\
  &            & T0+24897.0:T0+25145.9  & 12.93$\pm$0.01 \\
  &            & T0+36759.4:T0+37018.0  & 12.80$\pm$0.01 \\
  & 	       & Q (UVOT Tool) & 13.2\\
$uvw1$ & Swift/UVOT & T0+13822.0:T0+14383.7  & 13.75$\pm$0.01\\
     &            & T0+23383.6:T0+24283.3  & 13.59$\pm$0.01\\
     &            & T0+35853.2:T0+36752.9  & 13.36$\pm$0.01\\
     &            & Q (290 s)\tablenotemark{b}			  & 14.03$\pm$0.01\\
$uvm2$ & Swift/UVOT & T0+12915.4:T0+13815.1  & 14.86$\pm$0.02 \\
     &            &T0+30983.6:T0+31047.9  & 14.76$\pm$0.05 \\
     &            &T0+34946.5:T0+35846.2  & 14.05$\pm$0.01 \\
     &		  &T0+42546.2:T0+43008.0  & 14.77$\pm$0.02 \\
     &            & Q (334 s)\tablenotemark{b}            & 15.27$\pm$0.03 \\
$uvw2$ & Swift/UVOT & T0+29164.3:T0+30064.1  & 14.90$\pm$0.01\\
     &            & T0+40727.6:T0+41627.4  & 14.89$\pm$0.01 \\
     &             & Q (334 s)\tablenotemark{b}           & 15.10$\pm$0.02 \\
r    & LT         &T0+900  (30s)\tablenotemark{c}              & 7.4$\pm$0.2 \\
\enddata
\tablenotetext{a}{Time in seconds since the BAT trigger}
\tablenotetext{b}{Observations were made almost one year after the flare, to determine the quiescent
magnitudes in these filters.  The time in parenthesis gives the exposure time for this observation.}
\tablenotetext{c}{Observations at the Liverpool Telescope consisted of 3 10 second exposures taken
roughly 15 minutes after the trigger.}
\end{deluxetable}

\begin{deluxetable}{lllllll}
\tablewidth{0pt}
\tablenum{2}
\tablecolumns{6}
\tablecaption{BAT Spectral Fit Results\label{tbl:batspec}\tablenotemark{a}}
\tablehead{\colhead{Time Interval\tablenotemark{b}} & \colhead{Temp.} & \colhead{VEM} 
& \colhead{$\chi^{2,}$\tablenotemark{c}}  & \colhead{f$_{X}$ (14-100 keV)}  & \colhead{f$_{X}$/f$_{\rm bol}$}  \\
\colhead{(s)} & \colhead{(keV)} & \colhead{(10$^{54}$ cm$^{-3}$)} &\colhead{} & \colhead{(10$^{-9}$ erg cm$^{-2}$ s$^{-1}$)} & \colhead{} 
}
\startdata
T0-39: T0  & 12.0$^{+3.8}_{-2.7}$& 6.3$^{+3.5}_{+-2.1}$&46.87 &11.7$^{+0.7}_{-2.05}$& 0.62\\
T0: T0+112  & 11.7$^{+2.7}_{-2.2}$ & 4.7$^{+2.3}_{-1.4}$&35.36  &8.5$^{+0.2}_{-0.8}$ & 0.45\\
T0+112: T0+165.6 & 8.9$^{+2.7}_{-1.9}$& 4.8$^{+3.6}_{-2.0}$&24.45 &5.0$^{+0.2}_{-1.0}$ & 0.26\\
T0+165.6: T0+961 & 10.0$^{+5.8}_{-2.8}$& 0.81$^{+0.81}_{-0.41}$&34.91 &1.10$^{+0.04}_{-0.29}$& 0.06\\
\enddata
\tablenotetext{a}{Global metallicity $Z$ fixed to 0.4.}
\tablenotetext{b}{Time in seconds since the BAT trigger}
\tablenotetext{c}{37 degrees of freedom for each spectrum}
\end{deluxetable}

\begin{deluxetable}{clllllll}
\rotate
\tablewidth{0pt}
\tablenum{3}
\tablecolumns{8}
\tablecaption{XRT Spectral Fit Results \label{tbl:xrtspec}}
\tablehead{\colhead{Time Interval\tablenotemark{a}} & \colhead{kT$_{1}$} & \colhead{VEM$_{1}$} & \colhead{kT$_{2}$} & \colhead{VEM$_{2}$}
& \colhead{Z} & 
\colhead{$\chi^2$,d.o.f.} &
\colhead{f$_{X}$ (0.3-10 keV)} \\
\colhead{(s)} & \colhead{(keV)} & \colhead{(10$^{54}$ cm$^{-3}$)} & \colhead{(keV)} & \colhead{(10$^{54}$ cm$^{-3}$)} & \colhead{} & 
 \colhead{}  &
\colhead{(10$^{-9}$ erg cm$^{-2}$ s$^{-1}$) }
}
\startdata
T0+171:T0+306  &  $1.196^{+0.083}_{-0.184}$ &  0.58$^{+0.28}_{-0.18}$  &
$8.383^{+1.225}_{-0.843}$ & $3.69^{+0.14}_{-0.13}$  &
$0.477^{+0.110}_{-0.099}$  &  
  354.2,375 &
$24.5^{+0.5}_{-0.6}$   \\

T0+306:T0+508  &  $0.990^{+0.036}_{-0.041}$  & $0.36^{+0.08}_{-0.09}$ &
$7.068^{+0.747}_{-0.634}$ & $2.35^{+0.06}_{-0.06}$ & $0.562^{+0.119}_{-0.109}$
&     406.2,354 
 & $15.8^{+0.2}_{-0.4}$   \\

T0+508:T0+671  &  $0.956^{+0.053}_{-0.066}$ & $0.32^{+0.12}_{-0.08}$  &
$6.036^{+0.781}_{-0.656}$ & $1.74^{+0.07}_{-0.06}$  &
$0.442^{+0.129}_{-0.113}$ &   
       307.5,281 &
$11.1^{+0.2}_{-0.3}$  \\

T0+671:T0+874  &  $0.789^{+0.026}_{-0.026}$  & $0.25^{+0.05}_{-0.04}$ &
$5.222^{+0.445}_{-0.303}$ & $1.51^{+0.04}_{-0.04}$ &
$0.439^{+0.085}_{-0.078}$ &  
      388.1,336 &
$9.2\pm 0.2$  \\

T0+874:T0+1118  &  $0.789^{+0.024}_{-0.025}$ & $0.247^{+0.052}_{-0.008}$   &
$4.828^{+0.343}_{-0.318}$ & $1.23^{+0.03}_{-0.03}$  &  $0.399^{+0.076}_{-0.069}$ &
       429.6,324  &
$7.4\pm 0.1$   \\

T0+1118:T0+1451  &  $0.759^{+0.024}_{-0.025}$ &  $0.18^{+0.04}_{-0.03}$   &
$3.951^{+0.221}_{-0.205}$ & $0.91^{+0.02}_{-0.02}$ &
$0.414^{+0.075}_{-0.069}$ &   
310.3,310 &
$5.2\pm0.1$    \\   

T0+1451:T0+1971  &  $0.772^{+0.018}_{-0.019}$  & $0.16^{+0.03}_{-0.02}$   &
$3.754^{+0.222}_{-0.207}$ & $0.55^{+0.02}_{-0.02}$  &
 $0.429^{+0.069}_{-0.063}$ &   
     366.3,300  &
$3.3\pm0.1$ \\

T0+1971:T0+2215  &  $0.759^{+0.021}_{-0.022}$  & $0.095^{+0.02}_{-0.01}$   &
$3.271^{+0.147}_{-0.139}$  & $0.40^{+0.01}_{-0.01}$ &  $0.432^{+0.064}_{-0.059}$ &
   304.2,292 & 
$2.22\pm 0.03$\\

T0+2215:T0+2517  &  $0.771^{+0.021}_{-0.022}$ &  $0.065^{+0.010}_{-0.008}$  &
$3.140^{+0.136}_{-0.131}$ & $0.322^{+0.010}_{-0.009}$ &
$0.497^{+0.065}_{-0.061}$ &   
365.4,293 & 
$1.77\pm 0.03$ \\ 

T0+2517:T0+2798  & $0.773^{+0.025}_{-0.027}$ & $0.042^{+0.007}_{-0.006}$ &
 $2.811^{+0.137}_{-0.146}$ & $0.258^{+0.010}_{-0.010}$ &
 $0.589^{+0.083}_{-0.076}$& 
 271.8,259 &
$1.39^{+0.02}_{-0.03}$ \\ 

T0+6160:T0+4.3$\times 10^4$  & $0.727^{+0.025}_{-0.027}$ & $0.0052^{+0.0009}_{-0.0006}$   &   $2.107^{+0.062}_{-0.053}$
& $0.0096^{+0.0006}_{-0.0006}$  & $0.382^{+0.062}_{-0.053}$ &
236.3,175 &
$0.057^{+0.001}_{-0.001}$ \\

\enddata
\tablenotetext{a}{Time in seconds since the BAT trigger}
\end{deluxetable}

\begin{deluxetable}{cllllll}
\tablewidth{0pt}
\tablenum{4}
\tablecolumns{7}
\tablecaption{Time-Resolved Equivalent Widths of the Fe 6.4 keV line\label{tbl:fek}}
\tablehead{\colhead{Time Interval\tablenotemark{a}} & \colhead{E\tablenotemark{b}} & \colhead{EW\tablenotemark{b}} & \colhead{F\tablenotemark{b}} & \colhead{$\Delta \chi^{2}$/dof}
&\colhead{P$_{F-test}$} & \colhead{P$_{MC}$} \\
\colhead{(s)} & \colhead{(keV)} & \colhead{(eV)} & \colhead{(10$^{-3}$ photons cm$^{-2}$ s$^{-1}$)} & \colhead{}
&\colhead{(\%)} & \colhead{(\%)}
}
\startdata
T0+171:T0+306  & $6.44\pm 0.08$    & $146^{+67}_{-72.5}$ & $27.3^{+14.7}_{-12.9}$  &
10.7/2  & $99.672$ & $99.94$ \\
T0+306:T0+508  & $6.42^{+0.31}_{-0.41}$   & 
$52.2^{+64.8}_{-52.2}$ & $6.96^{+8.23}_{-6.96}$  & 1.8/2 & $54.200$ & 55.42\\
T0+508:T0+671  & $6.41\pm 0.04$   &  $320\pm 102$ & $26.4^{+9.6}_{-9.5}$ & 20.7/2 & $99.994$ & 99.97\\
T0+671:T0+874  & $6.42^{+0.07}_{-0.10}$   &  $103^{+95}_{-81}$  &
$6.51^{+4.67}_{-4.37}$ & 5.2/2 & $89.500$ & 99.19 \\
T0+874:T0+1118  & $6.46^{+0.05}_{-0.06}$   & $143^{+106}_{-114}$ &
$6.86^{+4.18}_{-4.04}$ & 7.7/2 & $94.563$ & 99.92\\
T0+1118:T0+1451  & $6.41^{+0.15}_{-0.41}$   & $84.3^{+155.7}_{-84.3}$ &
$2.65^{+2.35}_{-2.33}$  & 3.5/2 & $82.600$ & 97.69\\
T0+1451:T0+1971  & $6.45^{+0.10}_{-0.09}$   &   $91^{+139}_{-91}$ &
$2.12^{+2.02}_{-1.63}$ &  4.5/2 &  $84.100$ &ND\tablenotemark{c} \\
T0+1971:T0+2215  & $6.38^{+0.11}_{-0.12}$   & $94.8^{+288.2}_{-94.8}$ &
$1.20^{+0.94}_{-1.06}$ & 3.4/2 & $80.400$ & ND\tablenotemark{c}\\
T0+2215:T0+2517  & $6.11^{+0.20}_{-0.11}$   & $331^{+52}_{-281.3}$ & $1.65\pm 0.85$ &
10.4/2 & $98.502$ & 97.30\\
T0+2517:T0+2798  & $6.22^{+0.13}_{-0.22}$   & $295^{+302}_{-295}$ &
$0.95^{+0.77}_{-0.71}$  & 4.8/2 & $89.900$ & ND\tablenotemark{c}\\
\enddata
\tablenotetext{a}{Time in seconds since the BAT trigger}
\tablenotetext{b}{Errors are 3$\sigma$}
\tablenotetext{c}{No Data -- Monte Carlo calculations not done for this time interval due to low significance from F-test
(see \S 3.2.2 for details)}
\end{deluxetable}

\begin{deluxetable}{lcc}
\tablewidth{0pt}
\tablenum{5}
\tablecolumns{3}
\tablecaption{Spectral Fits to Time Interval T0+171:T0+961 s \label{tbl:xrtbat}}
\tablehead{\colhead{} & \colhead{$2T$} & \colhead{$3T$} }
\startdata
\hline
\multicolumn{3}{c}{ --- XRT+BAT --- } \\
\hline
kT$_1$ (keV)&0.98$^{+0.02}_{-0.03}$ &0.66$^{+0.08}_{-0.07}$ \\
VEM$_1$ (10$^{54}$ cm$^{-3}$) & 0.35$^{+0.05}_{-0.04}$&0.11$^{+0.04}_{-0.02}$ \\
kT$_2$ (keV)&6.68$^{+0.37}_{-0.24}$ &1.24$^{+0.09}_{-0.09}$ \\
VEM$_2$ (10$^{54}$ cm$^{-3}$)&2.10$^{+0.03}_{-0.03}$ &0.32$^{+0.04}_{-0.04}$ \\
kT$_3$ (kev) &-- &7.34$^{+0.31}_{-0.30}$ \\
VEM$_3$ (10$^{54}$ cm$^{-3}$)&-- &1.96$^{+0.05}_{-0.04}$ \\
A &0.43$^{+0.05}_{-0.05}$ &0.50$^{+0.05}_{-0.04}$ \\
E (keV)&6.43$^{+0.04}_{-0.03}$ &6.43$^{+0.23}_{-0.06}$ \\
N (photons cm$^{-2}$ s$^{-1}$)&(1.35$^{+0.35}_{-0.38}$)$\times$10$^{-2}$ &(1.31$^{+0.38}_{-0.39}$)$\times$10$^{-2}$ \\
$\chi^{2}$, dof &777.00, 589 &698.94, 587 \\
f(0.3--10 keV) (erg cm$^{-2}$ s$^{-1}$) & (1.390$^{+0.006}_{-0.011}$)$\times$10$^{-8}$ & (1.395$^{+0.008}_{-0.007}$)
$\times$10$^{-8}$\\
f(14--100 keV) (erg cm$^{-2}$ s$^{-1}$) & (1.04$^{+0.04}_{-0.05}$)$\times$10$^{-9}$ & (1.29$^{+0.06}_{-0.09}$)
$\times$10$^{-9}$ \\
\hline
\multicolumn{3}{c}{ --- XRT only ---} \\
\hline
kT$_1$ (keV)&0.98$^{+0.02}_{-0.03}$ &0.67$^{+0.09}_{-0.05}$ \\
VEM$_1$ (10$^{54}$ cm$^{-3}$) &0.35$^{+0.06}_{-0.04}$ &0.12$^{+0.06}_{-0.02}$ \\
kT$_2$ (keV)&6.67$^{+0.59}_{-0.28}$  &1.30$^{+0.2}_{-0.08}$ \\
VEM$_2$ (10$^{54}$ cm$^{-3}$)&2.10$^{+0.03}_{-0.03}$ &0.37$^{+0.05}_{-0.04}$ \\
kT$_3$ (kev) &-- &8.22$^{+0.76}_{-0.52}$ \\
VEM$_3$ (10$^{54}$ cm$^{-3}$)&-- &1.92$^{+0.05}_{-0.08}$ \\
A &0.43$\pm$0.05 &0.50$\pm$0.05\\
E (keV)&6.43$^{+0.07}_{-0.03}$ &6.44$\pm$0.04 \\
N (photons cm$^{-2}$ s$^{-1}$)&(1.35$^{+0.35}_{-0.39}$)$\times$10$^{-2}$ & (1.23$^{+0.37}_{-0.39}$)$\times$10$^{-2}$\\
$\chi^{2}$, dof &737.25, 550 & 646.69, 548 \\
f(0.3--10 keV) (erg cm$^{-2}$ s$^{-1}$) & (1.390$^{+0.008}_{-0.006}$)$\times$10$^{-8}$ & (1.410$\pm$0.008)$\times$10$^{-8}$ \\
\enddata
\end{deluxetable}

\begin{deluxetable}{ccc}
\tablewidth{0pt}
\tablenum{6}
\tablecolumns{3}
\tablecaption{Flare Radiated Energy Estimates \label{tbl:energy}}
\tablehead{\colhead{Energy Range} & \colhead{Time Range\tablenotemark{a}} & \colhead{Energy} \\
\colhead{(keV)} & \colhead{(s) } & \colhead{(10$^{33}$ erg)}
}
\startdata
14-100 & T0-39:T0+961  & 7.6\tablenotemark{b} \\
0.3-10 &  T0-39:T0+165.5  &  20\tablenotemark{c}\\
0.3-10 & T0+171:T0+19792 & 55\tablenotemark{d}\\
0.3-100 & T0+171:T0+961  & 37\tablenotemark{e}\\
0.3-10 & T0+171:T0+8300 & 58\tablenotemark{f}\\
\enddata
\tablenotetext{a}{Time in seconds since the BAT trigger}
\tablenotetext{b}{Determined from fits to BAT spectra}
\tablenotetext{c}{Determined by extrapolating fits to BAT spectra in this energy range}
\tablenotetext{d}{Determined from fits to XRT spectra}
\tablenotetext{e}{Determined from joint fit to XRT and BAT spectra}
\tablenotetext{f}{Determined from exponential fit to flux vs time with a time constant of 647 s before T0+1000s,
and a time constant of 1042 s from T0+1000s:T0+8300s}
\end{deluxetable}

\clearpage

\end{document}